\documentclass[twoside,twocolumn,9pt]{article}
\usepackage[utf8]{inputenc}
\usepackage[T1]{fontenc}
\usepackage{extsizes}
\usepackage[super,sort&compress,comma]{natbib} 
\usepackage[version=3]{mhchem}
\usepackage[left=1.5cm, right=1.5cm, top=1.785cm, bottom=2.0cm]{geometry}
\usepackage{balance}
\usepackage{mathptmx}
\usepackage{sectsty}
\usepackage{graphicx} 
\usepackage{lastpage}
\usepackage[format=plain,justification=justified,singlelinecheck=false,font={stretch=1.125,small,sf},labelfont=bf,labelsep=space]{caption}
\usepackage{float}
\usepackage{fancyhdr}
\usepackage{fnpos}
\usepackage[english]{babel}
\addto{\captionsenglish}{%
  
}
\usepackage{array}
\usepackage[usenames,dvipsnames]{xcolor}
\usepackage[colorinlistoftodos]{todonotes}
\usepackage{droidsans}
\usepackage{charter}
\usepackage[T1]{fontenc}
\usepackage{setspace}
\usepackage[compact]{titlesec}
\usepackage{hyperref}

\usepackage{epstopdf}

\definecolor{cream}{RGB}{222,217,201}

\begin{document}



\setcounter{secnumdepth}{5}

  \noindent\LARGE{\textbf {Doping isolated one-dimensional antiferromagnetic semiconductor Vanadium tetrasulfide (\ce{VS4}) nanowires with carriers induces half-metallicity $^\dag$}} \\

  \noindent\large{Shuo Li \textit{$^{a}$}, Junjie He \textit{$^{a,b}$}, Petr Nachtigall \textit{$^{a}$}, Luk\'a\v s Grajciar \textit{$^{a}$},  Federico Brivio \textit{$^{a}$}} \\

 \noindent\normalsize{Quasi one-dimensional (1D) vanadium tetrasulfide (\ce{VS4}) nanowires (NWs) are synthetic semiconductors which combine with each other through Van der Waals interactions to form bulk phases. However, the properties of these individual nanowires remain unknown. Nevertheless, our calculations of their stability indicate that \ce{VS4} NWs can be separated from their bulk structures. Accordingly, we theoretically investigated the geometrical, electronic, and magnetic properties of bulk phase and isolated \ce{VS4} NWs. 
Our results indicate that both bulk phase and isolated \ce{VS4} NWs are semiconductors with band gaps of 2.24 and 2.64 $eV$, respectively, and that they prefer the antiferromagnetic (AFM) ground state based on DFT calculations. These calculations also suggested that isolated \ce{VS4} NWs show half-metallic antiferromagnetism upon electron and hole doping because carrier doping splits the spin degeneracy to induce local spin polarisation. As a result,
spin polarisation currents in isolated \ce{VS4} NWs can be manipulated with locally applied gate voltage. 
Therefore, these 1D AFM materials have a high potential for advancing both fundamental research and spintronic applications because they are more resistant to magnetic perturbation than their 1D ferromagnetic counterparts.
 
} 


\renewcommand*\rmdefault{bch}\normalfont\upshape
\rmfamily
\section*{}
\vspace{-1cm}


\footnotetext{\textit{$^{a}$Department of Physical and Macromolecular Chemistry \& Charles University Center of Advanced Materials, Faculty of Science, Charles University in Prague, Hlavova 8, 128 43 Prague 2, Czech Republic; E-mail: briviof@natur.cuni.cz}}
\footnotetext{\textit{$^{b}$Bremen Center for Computational Materials Science, University of Bremen, Am Fallturm 1, 28359 Bremen, Germany}}



\section{Introduction}
Research aimed at improving the performance of electronics has pushed the physical limits of these devices.
However, further advances in information technology require developing alternatives to electronics\cite{del2011nanometre, tomioka2012iii, ferain2011multigate, mack2011fifty}. For this purpose, many new methodologies have been proposed, such as molecular electronics, nanoelectronics\cite{lu2010nanoelectronics, akinwande2014two, berger2004ultrathin}, spintronics\cite{wolf2001spintronics, felser2007spintronics, li2016first}, and valleytronics\cite{schaibley2016valleytronics, zhong2017d,jungwirth2018multiple}. Among these new developments, spintronics stands out for its compatibility with conventional electronics. Consequently, spintronics can be used to broaden the possibilities of conventional electronics \cite{wolf2001spintronics, vzutic2004spintronics}. In contrast to electronics, in which electrical charges are manipulated to induce a current, spintronics aims to exploit the two spin polarization of unpaired electrons to create a spin-polarised current, thus enhancing the performance of semiconductors, in terms of electrical conductivity and transport of binary information. Accordingly, spintronics enables us to improve the design of logic components, including memory boards and transistors\cite{chumak2015magnon,camsari2015modular}.

Spintronics is based on the ability to control the intrinsic spin of the electron, which was first discovered\cite{mott1936resistance} in ferromagnetic (FM) metals but has since been proved viable in other classes of FM materials, most notably in half-metallic ferromagnets\cite{picozzi2014ferroelectric}.
Unlike paramagnetic metals, FM metals have a different density of states for the two spins. Consequently, the latter have a neat magnetic momentum and exhibit a spin-polarised current at adequately low temperatures\cite{mott1936resistance, mott1936electrical}. Similarly, half-metallic ferromagnets have two spin channels (which guarantees a neat magnetic momentum), but only one of spin channels displays metallic character, whilst the other has a band-gap\cite{li2017low}. FM materials with non-null magnetic polarisation have stray fields, which can induce interference between different elements, thus limiting the down-scaling of devices\cite{li2016first,jungwirth2016antiferromagnetic, baltz2018antiferromagnetic, macdonald2011antiferromagnetic}). For this reason, most studies have focused on antiferromagnetic (AFM) materials. 

AFM materials have been primarily manipulated for spintronics by inducing half-metallic antiferromagnetism (HMAF)\cite{hu2012half, he2015prediction, nie2008possible}. HMAF materials were first proposed by H. van Leuken and R. A. de Groot\cite{van1995half}, who showed that many Heusler compounds, more specifically \ce{CrMnSb}, could be fully spin-compensated half-metallic materials. Heusler compounds are ternary compounds with two different magnetic centres at different sub-lattices, thus decoupling electronic from magnetic properties. Recently, these theoretical predictions have been experimentally confirmed in similar compounds, namely Half-Heusler: \ce{Mn2RuxGa} and \ce{Mn2Pt_xGa} alloys\cite{nayak2015design, kurt2014cubic}. 
These studies have mostly focused on bulk phases, while low-dimensional materials remain unexplored because their lower dimensionality requires different strategies to remove the spin degeneracy, either applying a bias voltage or using organic-inorganic materials, for example.
Gong et al. showed that the bilayer \ce{2H-VSe2} becomes HMAF when applying proper electric fields \cite{gong2018electrically}, while Ai \textit{et al.} designed a two-dimensional (2D) metal-organic HMAF (CoFePz), which paved the way to the development of organic HMAF\cite{ai2018two}.
Among its potential applications of HMAF,spin field-effect transistors (FET)\cite{chuang2015all, gong2018electrically} stand out for their ability to control the spin current\cite{deng2018gate}.
Moreover, carrier doping enables us to manipulate the electronic and magnetic properties of low-dimensional materials, which are theoretically and experimentally accessible\cite{li2014half, yuan2009high, he2019cr}.
Therefore, carrier doping is an effective strategy to control the spin current in low-dimensional magnetic materials.
Few materials, for example \ce{NbSe3}\cite{pham2018torsional}, can be isolated as true one-dimensional (1D) materials, which are joined in quasi-1D materials through Van der Waals forces.
Some 1D compounds have been proposed as candidates for spintronic applications, including metal trihydride molecular nanowires\cite{li2017half} (NWs), 1D metal benzenetetramine coordination polymers\cite{wan2018ambipolar}, Co-dithiolene molecular wires\cite{zhang2016electron}, transition metal tribromide NWs\cite{li2018robust}, transition metal trichalcogenide NWs\cite{pham2018torsional, ye2017amorphous}, and transition metal chalcogenide NWs\cite{shang2020atomic}. 
These materials have been theoretically investigated for their electronic and magnetic properties, but not as much  for their spintronic properties, due to the lack of experimental reports.

The Vanadium tetrasulfide (\ce{VS4}) is found in nature as a mineral, and was discovered in 1906\cite{hillebrand1907vanadium}.
Its linear chain-like structure is composed of two \ce{S2^{2-}} moieties connecting \ce{V^{4+}} centers.
The different chains are bound together by Van der Waals forces to form nano-rods in quasi-1D compounds\cite{flores2018beyond, zhou2016conductive, lui2015synthesis}.
\ce{VS4} NWs have been studied for other applications, such as batteries, capacitors, and photocatalysts\cite{flores2018beyond, rout2013synthesis, sun2015vanadium, wang2018graphene, wang2018highly}.
The possible oxidation states of Vanadium ions induce different magnetic properties that have been demonstrated in different materials: Vanadium dichalcogenides (\ce{VX2}, X = S, Se) monolayers, MXenes (\ce{VX2}, X = C, N)\cite{gao2016monolayer, frey2019surface}, and Haeckelite \ce{VS2}\cite{ma2017two}.
However, the geometric, electronic and magnetic properties of \ce{VS4} as a 1D NW are not clearly understood yet.
This study aimed to assess how the magnetic properties of \ce{VS4} are affected by its dimensionality and how \ce{VS4} can be used for spintronic applications.

\section{Methods}
All calculations were performed within the density functional theory (DFT) as implemented in the Vienna \textit{ab-initio} simulation package (VASP)\cite{kresse1993g, kresse1999g}.
The structural properties have been determined using the Perdew Burke Ernzerhof (PBE) version of the generalised gradient approximation (GGA)\cite{perdew1996generalized}, while the electronic and magnetic properties have been calculated with the hybrid HSE06\cite{heyd2003hybrid} functional. 
The dispersion forces have been included in the DFT-D3 method \cite{grimme2010consistent}. 
An energy cut-off of 500 \ce{eV} was employed to define the plane-wave basis sets, considering the wavefunction converged for total energy variations below $10^{-6}$ \ce{eV}. The same parameters have been used for both PBE and HSE06 calculations.
The structures were fully optimised to minimise the forces below \ce{-0.02} \ce{eV} /{\AA}. For both geometry relaxation and electronic properties in the Brillouin zone, this material was sampled using a Monkhorst–Pack k-point mesh of $5\times5\times5$ and $1\times1\times5$ for the bulk phase and isolated NWs, respectively.
The isolated NW was simulated including a vacuum space of 15 {\AA} in the x and y directions.
The high symmetry points of the first Brillouin zone for the band structure of the \ce{VS4} bulk phase are shown in Figure S1. 
We calculated phonon spectra using the finite differences method, as implemented in VASP. Post-processing and analysis have been performed using the software PhonoPy\cite{togo2015phonopy}. 
For this task, we have used tighter convergence criteria, more specifically $10^{-8}$ \ce{eV} for the wavefunction and $-0.001$ \ce{eV}/{\AA} for the forces. In addition, we address the kinetic stability of the structures performing a set of \textit{ab-initio} molecular dynamics (AIMD) calculations, as implemented in VASP. These calculations have been completed using the Nos\'e algorithm\cite{nose1984unified} in the NVT ensemble at room temperature (300 K) for the duration of 9 ps.
The formation energy ($\ce{E_{form}}$) of each \ce{VS4} unit formula is calculated as:
\begin{equation}
E_{form} = E(VS_4)/n - E(V) - 4E(S)
\end{equation}
where $E(VS_4)$ is the total energy of the $\ce{VS4}$ bulk phase or the isolated NW containing $n$ unit formula, while $\ce{E[V]}$ and $\ce{E[S]}$ are the single atom energies from the bulk phases of cubic \ce{V} and \ce{S}\cite{watanabe1974crystal}.
Carrier doping was simulated by removing or adding electrons from the system with the homogeneous background charge to keep charge neutrality.
 
The magnetic properties have been analysed using a collinear model for simplicity. 
We considered FM and AFM states to calculate the preferred magnetic ground state structures of NW (Figure S2). To account for these effects, we used the Monte Carlo method to solve a simple Ising model\cite{mccoy2014two}:
\begin{equation}
H = -\sum_{ij} J_{intra}M_i\cdot M_j
\end{equation}
where $\ce{J_{intra}}$ is the nearest-neighbour exchange-coupling parameter of intra-NWs and $M$ is the spin magnetic moment per chemical formula. The $\ce{J_{intra}}$ value can be calculated by the exchange energy ($\ce{E_{ex}} = \ce{E_{FM}} - \ce{E_{AFM}}$). 
The exchange coupling parameters have been used to calculate the N\'eel temperature (\ce{T_N}) performing a Monte Carlo simulation on a 150 1D chain lattice with $10^{-5}$ steps for each temperature using the open-source software ALPS\cite{albuquerque2007alps}. 
\section{Results}
\subsection{Structural analysis}
\begin{figure}[h]   
\centering
  \includegraphics[height=6cm]{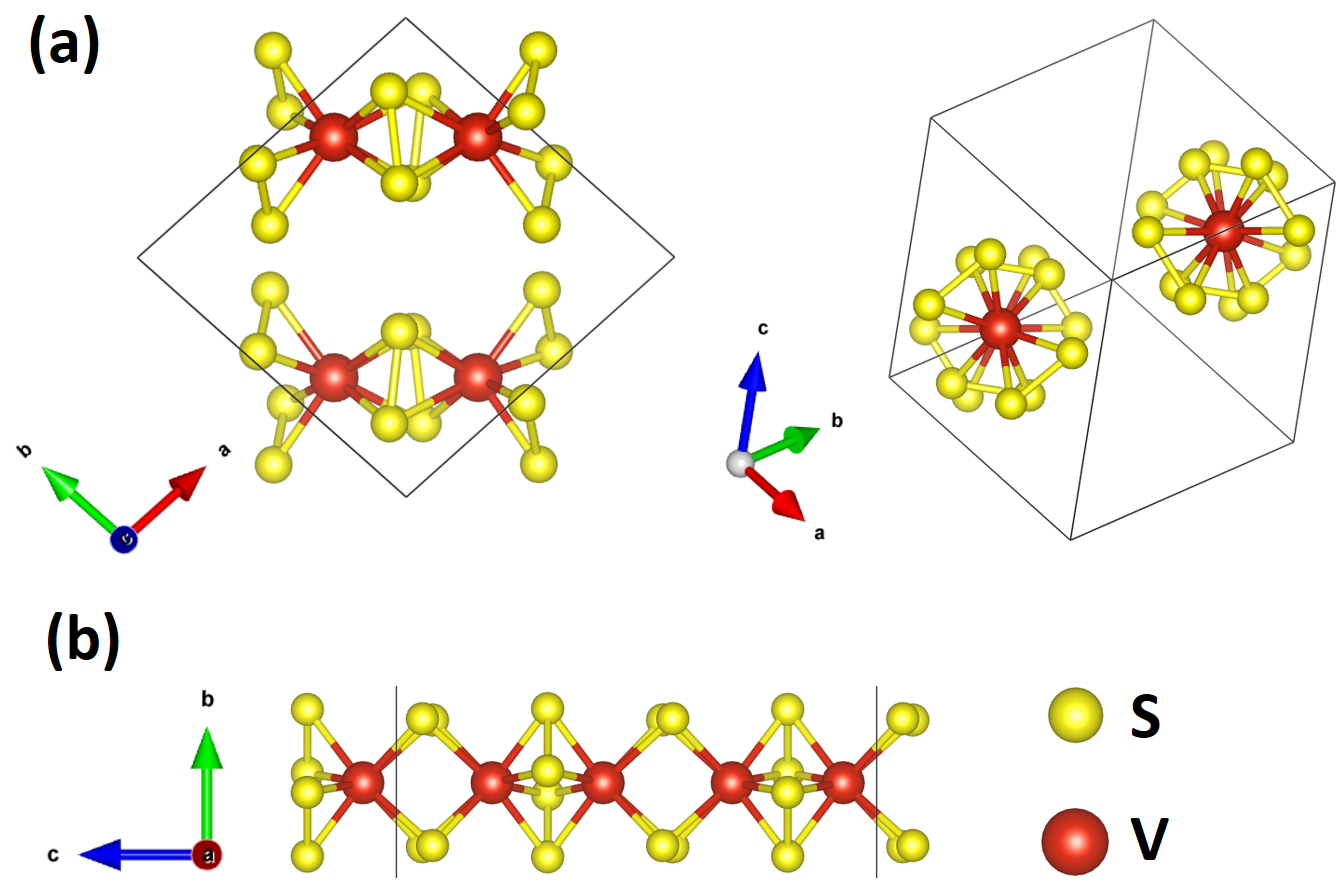} 
  \caption{(a) Primitive cell of the \ce{VS4} bulk phase from two directions, that is, along (001) direction, on the left panel, and along (111) direction, on the right panel. (b) Isolated \ce{VS4} NW. The repeating units are marked by the black lines.}
  \label{fig:structure}
\end{figure}

Figure \ref{fig:structure}a shows the primitive cell of \ce{VS4} bulk phase (C2/c \#15) in which each \ce{V} atom is coordinated with eight \ce{S} atoms forming four dimers.
From the bulk phase, we have built the isolated NW unite cell (Figure \ref{fig:structure}b). The geometric data on the \ce{VS4} bulk phase and on the isolated \ce{VS4} NW are summarised in Table 1. 
Both the bulk and isolate NW phases of \ce{VS4} are dynamically stable, as shown by the absence of an imaginary frequency in the phonon dispersion (figure S3a).
The stability of these materials is also confirmed by AIMD simulation at room temperature, thus suggesting that individual NWs can be isolated (figure S3b). 

\begin{table*}[]
\caption{Structural parameters, electronic and magnetic properties of the bulk phase and isolated NW for \ce{VS4}. \ce{E_{form}} is the formation energy ($eV$). \ce{L_{S-dimer}} is S dimer bond-length (\AA). \ce{L_{V-V}} is the V-V bond length. \ce{L_{V-S}} is the V-S bond length. \ce{E_g} is the band gap. \ce{E_{ex}} is the exchange energy (\ce{eV}). \ce{$\mu$_{B}} is the local magnetic moment of the V atoms. }
\centering
\begin{tabular}{ccccccccccc}
\hline
configurations & \ce{E_{form}}  & \ce{L_{S-dimer}} & \ce{L_{V-V}} & \ce{L_{V-S}} & \ce{E_g} & \ce{E_{ex}}  & \ce{$\mu$_{B}}   \\ \hline
Bulk     & -2.59     & 2.03        & 2.77, 3.20       & 2.54, 2.39         & 2.24      & 0.961    & 1.15       \\
NW      & -2.06     & 2.03        & 2.77, 3.22       & 2.55, 2.39         & 2.65      & 0.948    & 1.17        \\ \hline
\end{tabular}
\end{table*}
 
\subsection{Electronic and magnetic properties}
We now focus on the electronic and magnetic properties of the \ce{VS4} bulk phase and the isolated NW.
Both the bulk phase and the isolated NW have an AFM ground state, and their stability is given by \ce{E_{ex}}, as outlined in Table 1.
The magnetic momentum of \ce{V} is similar in the bulk phase and in the isolated NW, with values of 1.15  and 1.17 \ce{$\mu$_B}, respectively.
The bulk phase has an indirect band gap of 2.24 \ce{eV} between the \ce{Y} and \ce{Z} points of the Brillouin Zone.
The calculated electronic band gap largely overestimated the experimental optical band gap of \ce{VS4} films (approximately 1.35 \ce{eV})\cite{flores2018beyond} which is usually observed in similar cases\cite{ZHOU2018118,URREHMAN2017163,PingOrigin}. 
This difference of band gaps could be attributed to the experimental conditions in which the optoelectronic properties depend on the sulfur partial pressure of synthesis and on the morphology of the sample \cite{flores2018beyond}.
Most importantly, our model for the bulk phase assumes perfect crystallinity, which is unlikely to occur in the family of low-dimensional compounds.

\begin{figure}[h]   
\centering
  \includegraphics[height=6cm]{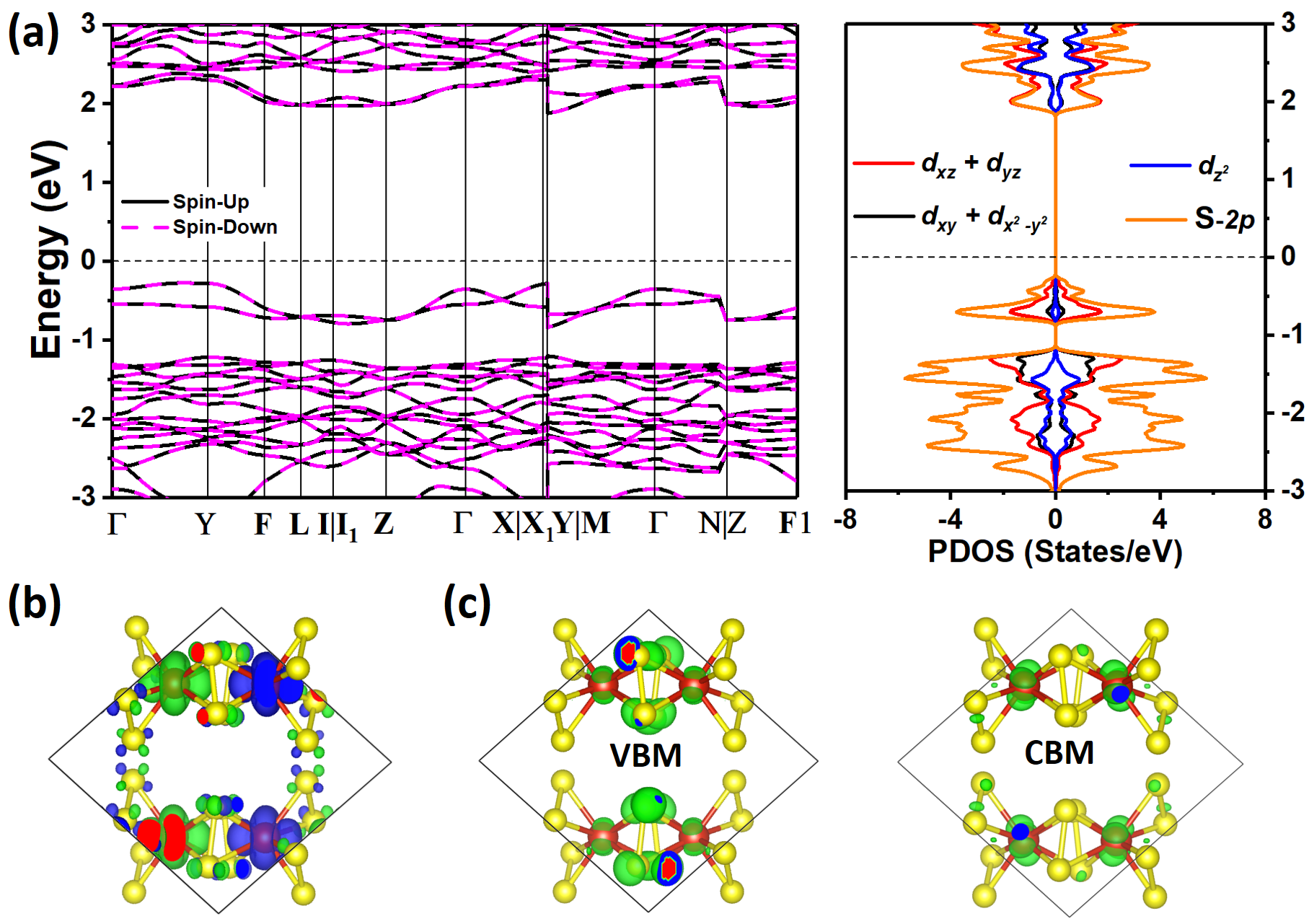} 
  \caption{(a) Band structures and PDOS and (b) spin-polarised density of the \ce{VS4} bulk phase, where spin-up and spin-down densities are shown in green and blue, respectively. (c) The valence band maximum (VBM) and the conduction band minimum (CBM) are shown in green. The isosurface is 0.005 \ce{e}  $\cdot$  \ce{Bohr^{-3}}.} 
  \label{fig:bulk}
\end{figure}

The spin density (Figure \ref{fig:bulk}b) shows the AFM G-type motif of the material\cite{takashiAtype}.
The valence band maximum (VBM) and the conduction band minimum (CBM) of the \ce{VS4} bulk phase are formed by the overlap of \ce{S} 2$p$ orbitals with \ce{V} $3d$ levels as shown in the partial density of states (PDOS) (Figure \ref{fig:bulk}a). 
The \ce{V} $3d$ orbitals split into a non-degenerate $\ce{d_{z^2}}$ orbital and into two 2-fold degenerate $\ce{d_{xz}}/\ce{d_{yz}}$ and $\ce{d_{xy}}/\ce{d_{x^2 -y^2}}$ orbitals. The local \ce{V} $3d$ orbitals can induce an antiparallel spin arrangement on neighboring \ce{V} via the double-exchange mechanism (Figure \ref{fig:bulk}b). The VBM and CBM of the \ce{VS4} bulk phase derive from the \ce{S} 2$p$ orbitals and \ce{V} $\ce{d_{xz}} + \ce{d_{yz}}$ orbitals (Figure \ref{fig:bulk}c), in line with its PDOS.

\begin{figure}[h]   
\centering
  \includegraphics[height=9cm]{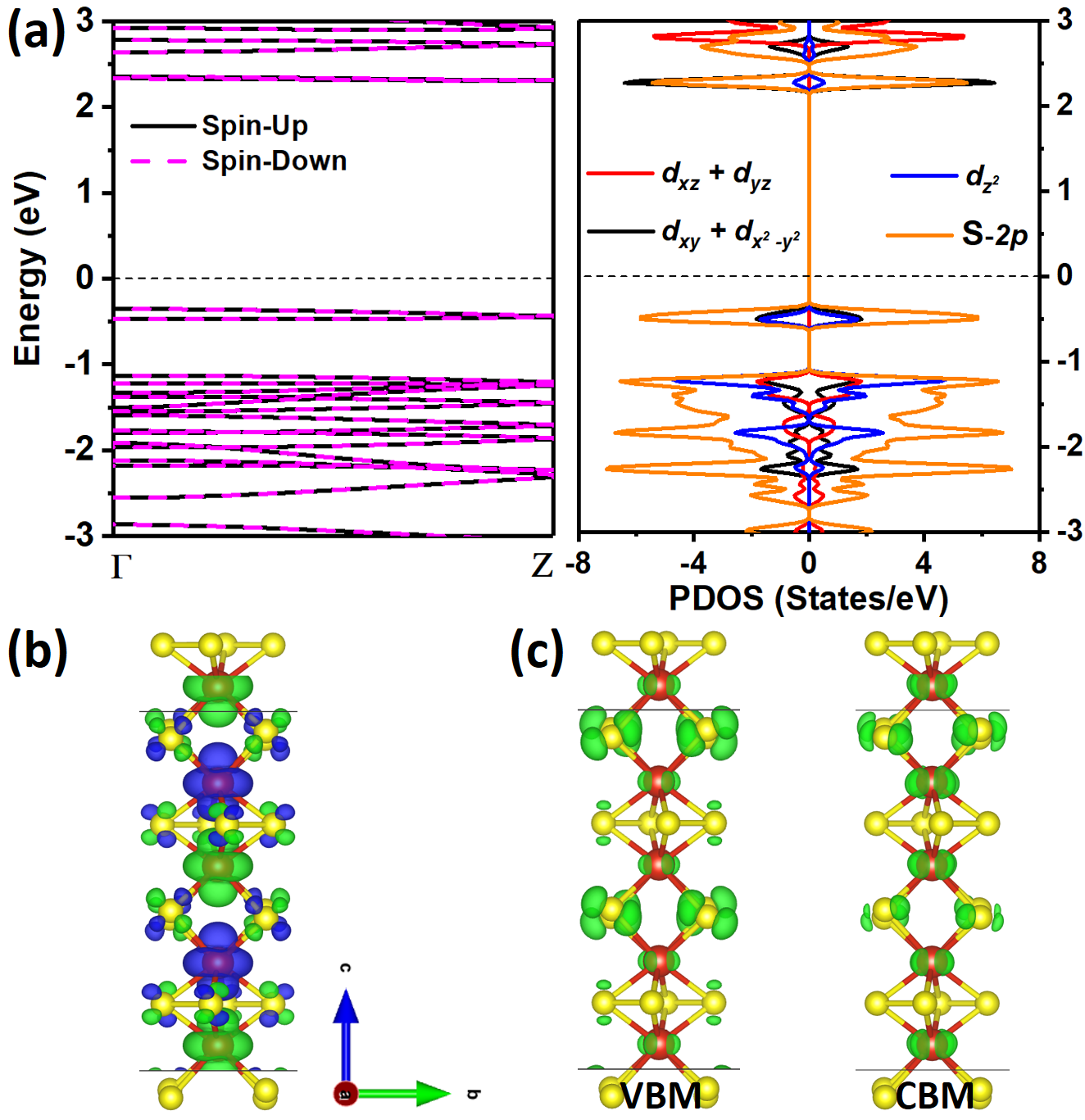} 
  \caption{(a) Band structures and PDOS and (b) spin polarised density of the isolated \ce{VS4} NW, where spin-up and spin-down densities are shown in green and blue, respectively. (c) VBM and CBM are shown in green. The isosurface is 0.005 \ce{e} $\cdot$ \ce{Bohr^{-3}}.}
  \label{fig:isolated}
\end{figure}

The electronic structure of the isolated \ce{VS4} NW reflects the lower dimensionality of the material, which induces a larger band-gap of 2.65 \ce{eV} with a higher density of state (Figure \ref{fig:isolated}a).
The lack of interactions between different chains of \ce{VS4} affects the nature of the bond. We can observe a larger contribution of VBM and CBM in the isolated \ce{VS4} NW due to the \ce{S} 2$p$ states (Figure \ref{fig:isolated}a).  
The VBM and the CBM of the \ce{V} $3d$ orbitals of the isolated NW are different from those of the bulk phase. \ce{V} $\ce{d_{xy}}/\ce{d_{x^2 -y^2}}$ orbitals of the isolated \ce{VS4} NW contribute to the VBM and CBM.

\begin{figure}[h]   
\centering
  \includegraphics[height=10cm]{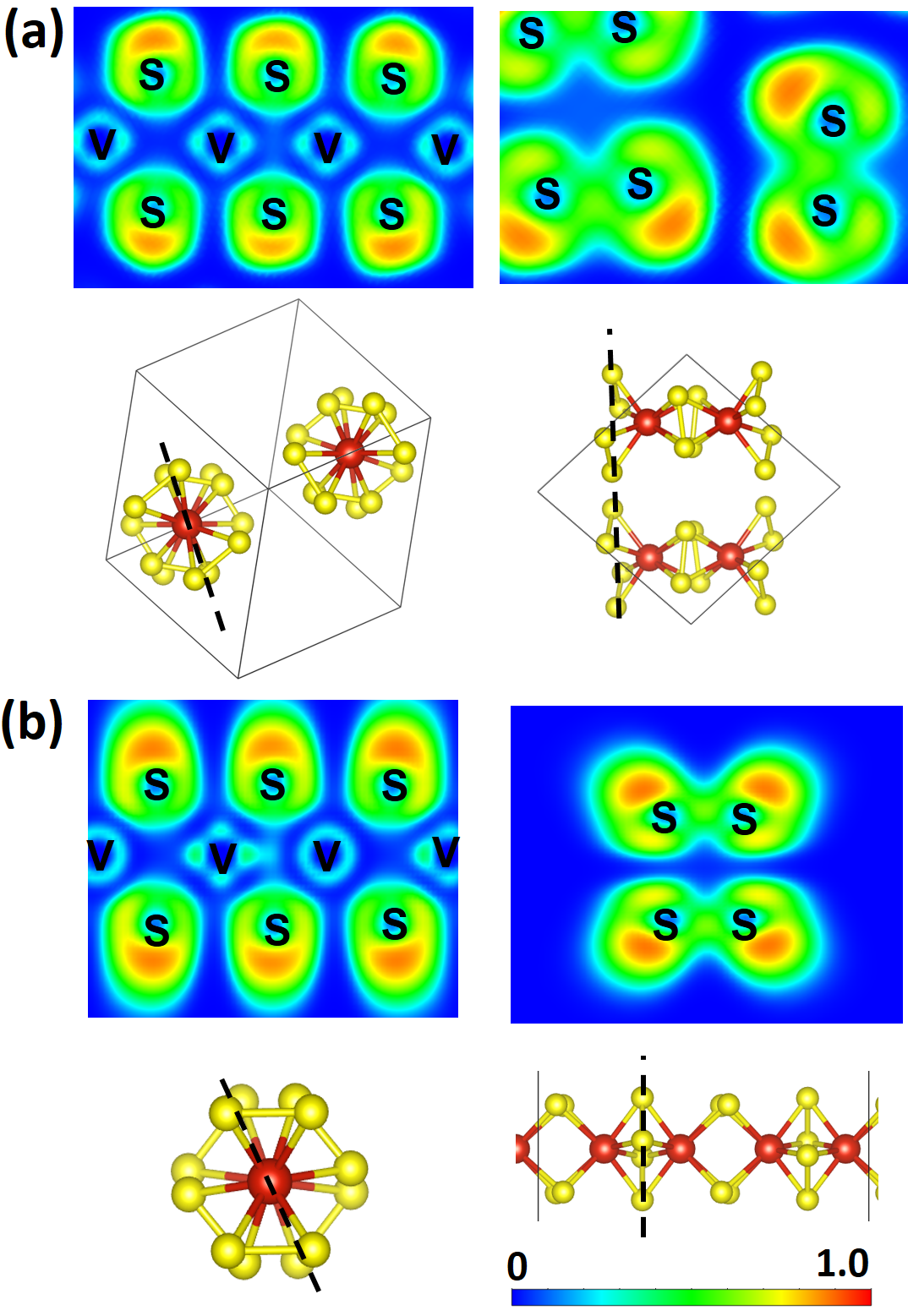} 
  \caption{Electron localisation function (ELF) maps of (a) the \ce{VS4} bulk phase and (b) isolated \ce{VS4} NW. The dot line is the chosen direction of the ELF slice. The unit of the color scale is "probability".}
  \label{fig:ELF}
\end{figure}

To further understand the change of electronic and magnetic properties between the \ce{VS4} bulk phase and the isolated \ce{VS4} NW, we plotted the electron localisation function (ELF) for the \ce{VS4} bulk phase and for the isolated \ce{VS4} NW (Figure \ref{fig:ELF}).
The ELF of \ce{S} atoms in the \ce{VS4} bulk phase is more localised than that of the isolated \ce{VS4} NW due to Van der Waals interactions of NWs in the \ce{VS4} bulk phase. 
Moreover, \ce{VS4} NW has distinct characteristics of localised $d$ electrons of \ce{V} atoms: more relatively itinerant \ce{V} $d$ electrons in the isolated \ce{VS4} NW than that in the \ce{VS4} bulk phase. 
Itinerant $d$ electrons of \ce{V} atoms can induce spin polarisation of neighbouring \ce{S} atoms, $e.g.$, an anti-parallel spin arrangement via a double-exchange mechanism. 
While the PDOS of $d$ orbitals of \ce{V} atoms differ between the \ce{VS4} bulk phase and the isolated \ce{VS4} NW. This difference in electronic properties does not affect the AFM character of these materials with a double-exchange mechanism, which also depends on the \ce{V} atoms.

\subsection{Macroscopic magnetic properties}
Magnetic order is more susceptible to temperature in materials with lower dimensionality\cite{Gambardellamagnetism} due to the lower number of interactions. As a result, magnetic fluctuation, which destroys the order, is more likely to happen in low-dimensional materials than in three-dimensional materials.
DFT calculations enable us to directly determine the magnitude of such interactions, but the value of \ce{T_N} cannot be simulated since it emerges as a statistic average. 
Nevertheless, we employed exchange interactions (obtained \textit{ab-initio}) to describe the Temperature dependence of our system using an Ising model solved with a Monte Carlo approach.
For the isolated \ce{VS4} NW, we obtained a \ce{T_N} of 210 $K$ (Figure S4).

\subsection{HMAF based FET}
\begin{figure}[h!]   
\centering
  \includegraphics[height=9.5cm]{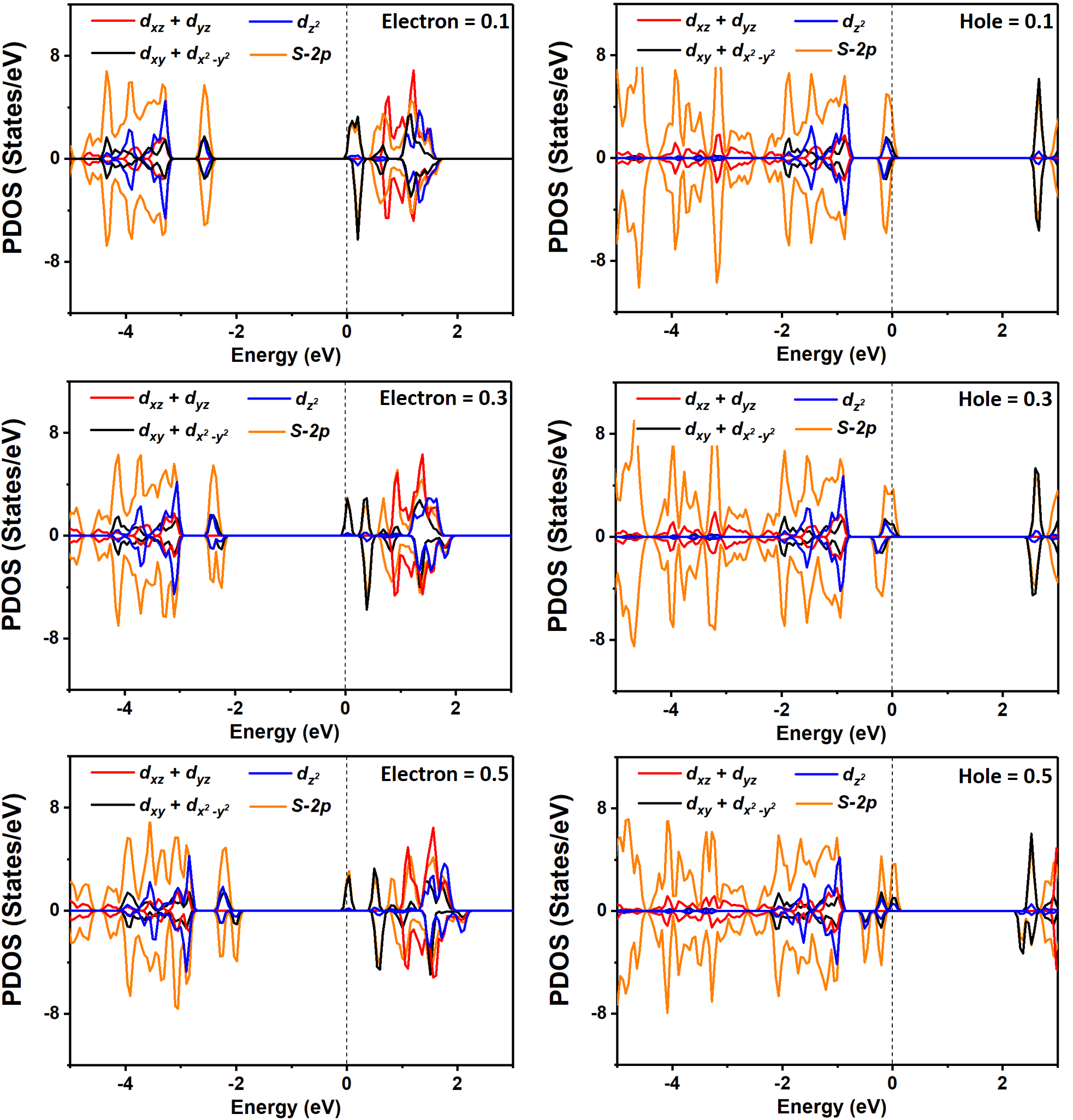} 
  \caption{PDOS of the isolated \ce{VS4} NW under carrier doping with a carrier concentration of 0.1, 0.3 and 0.5 electron (and hole) per unit cell.}
  \label{fig:carriers}
\end{figure}

Previous studies have shown that carrier doping manipulates spin currents through the voltage gate\cite{li2014half,he2019cr}.
We analysed doping concentrations of 0.1, 0.3, and 0.5 electron (and hole) per unit cell of the isolated \ce{VS4} NW (equivalent to 0.83, 2.50, and 4.17 $\times 10^{6} cm^{-1}$).
The isolated \ce{VS4} NW with carrier doping is HMAF, exhibiting complete spin-polarisation around the Fermi level (Figure \ref{fig:carriers}). 
Because carrier doping shifts the Fermi level and spin polarisation of \ce{S} and \ce{V} atoms, PDOS for the isolated \ce{VS4} NW with carrier doping show metallic states in the spin-up channel and semiconductor states (band gaps are over 2 $eV$.) in the spin-down channel. 
These effects are detected as small perturbations of the small AFM, which can be considered preserved for practical applications. 
The main contributions to the metallic states around the Fermi level derive from \ce{V} $\ce{d_{xy}}/\ce{d_{x^2 -y^2}}$ orbitals and \ce{S} 2$p$ orbitals.

\begin{figure}[h!]
\centering
\includegraphics[height=5.3cm]{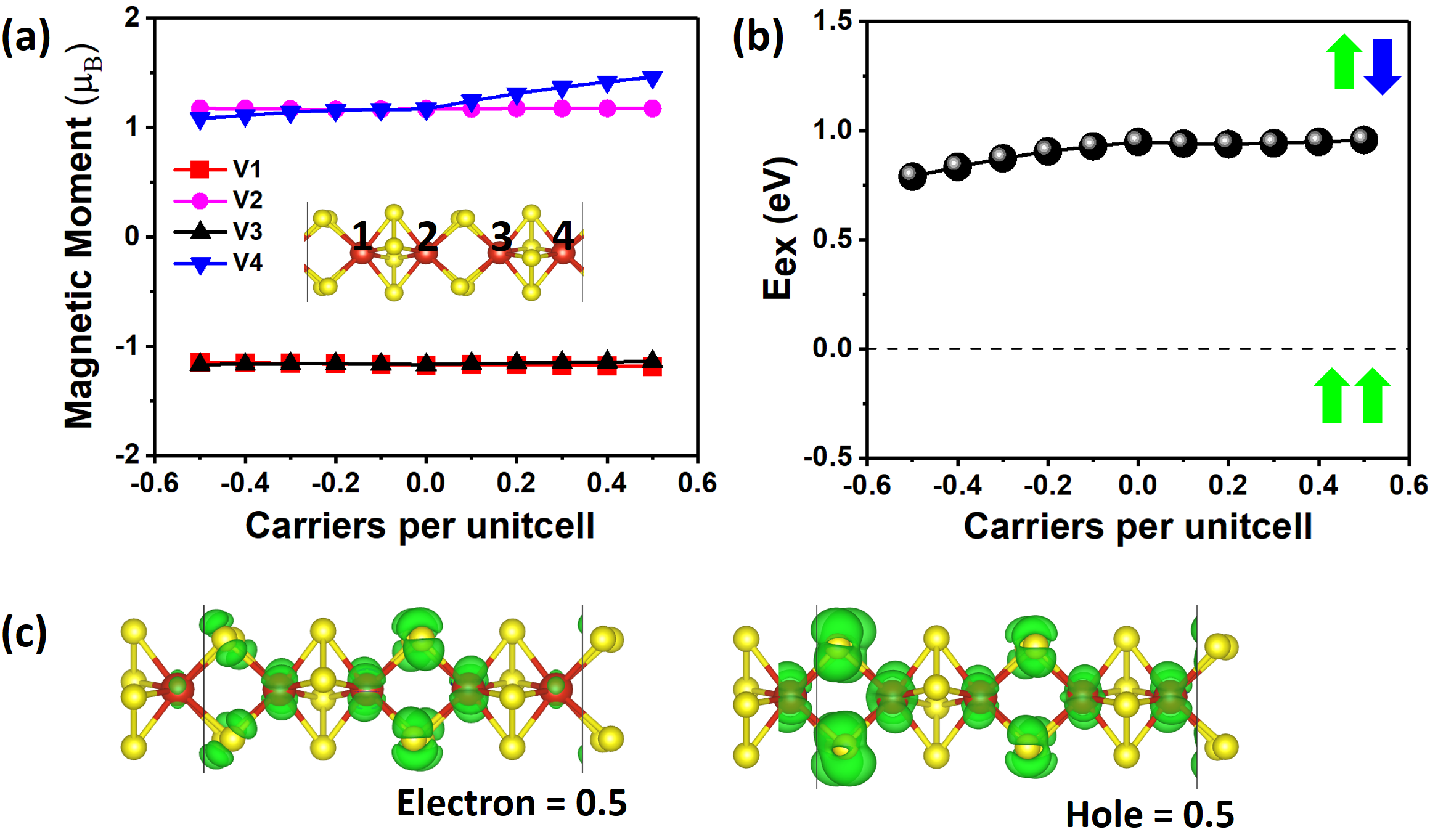}
\caption{(a) Local magnetic moment on each V atom of the isolated \ce{VS4} NW with carrier doping. (b) Exchange energy (\ce{E_{ex}}) under different carrier concentrations of the isolated \ce{VS4} NW. The positive and negative values refer to electron and hole doping, respectively. The up and down arrows indicate up- and down-spin, respectively. (c) partial charge density (around the Fermi level) of the isolated \ce{VS4} NW with 0.5 electron and hole doping.}
\label{fig:iband}
\end{figure}

The magnetic momentum of \ce{V} atoms is affected by carrier doping, and the magnetic momentum of each \ce{S} atom is very small. 
When doping with 0.5 electron, the magnetic momentum of one \ce{V} atom (marked 4) increases from 1.17 to 1.46 \ce{$\mu$_{B}} (Figure \ref{fig:iband}a). 
When doping with 0.5 hole, the magnetic momentum of one \ce{V} atom decreases from 1.17 to 1.08 \ce{$\mu$_{B}} (Figure \ref{fig:iband}a). 
The isolated \ce{VS4} NW retains its AFM state.
The \ce{E_{ex}} is rapidly decreased under hole doping and increased under electron doping. Our results indicate that the AFM state of the isolated \ce{VS4} NW is stable when injected with a low concentration of carriers.
Moreover, we visualise the partial charge density (around Fermi level) of the isolated \ce{VS4} NW at 0.5 electron and hole doping (Figure \ref{fig:iband}c).
The participation of \ce{V} and \ce{S} orbitals also differs: formed by \ce{V} $\ce{d_{xy}}/\ce{d_{x^2 -y^2}}$ and \ce{S} 2$p$ orbitals at 0.5 electron doping and by \ce{V} $\ce{d_{xy}}/\ce{d_{x^2 -y^2}}$, $\ce{d_{z^2}}$ and more \ce{S} 2$p$ orbitals at 0.5 hole doping.

\section{Discussion}
Our results indicate that the isolated \ce{VS4} NW is an AFM semiconductor with a high \ce{T_N} of 210 $K$.
This is an important result because ferromagnetic 1D materials, such as the quasi-1D organic ferromagnet (\ce{T_C} 0.56 $K$)\cite{takahashi1991discovery}, the tribromide NW of the half-metal \ce{VBr3} (\ce{T_C} 80 $K$)\cite{li2018robust}, the trihydride molecular NW of the half-metal \ce{CoH3} (\ce{T_C} 98 $K$)\cite{li2017half} and transition metal dihalide nanowires \ce{CuCl2}  (\ce{T_C} 14 $K$)\cite{tan2020one}, usually have much lower Curie temperatures (\ce{T_C}).
Furthermore, \ce{T_N} of the isolated \ce{VS4} NW is well above the temperature of liquid nitrogen (77K) and is also higher than that of different 2D antiferromagnets, such as bilayer \ce{CrI3} (\ce{T_N} 45$K$)\cite{huang2018electrical} or polyradical nanosheets (42.5$K$)\cite{yang2018antiferromagnetism}.
An even better prediction \ce{T_N} could be obtained for the isolated \ce{VS4} NW, but this predriction is a promising starting point for further experimental characterisation.

Moreover, the AFM character and large band gap of the isolated NWs prevent spin-polarised currents in \ce{VS4}. 
The local spin polarisation of ideal antiferromagnetism, which in turn allows spin-polarised currents, could be a technological solution\cite{he2019cr}.
Here, we consider a model where carriers are injected into our materials, a process commonly know as carrier doping in the literature, in analogy to chemical doping\cite{ueno2011discovery}.
A method for carrier doping consists of building a spin FET device (scheme shown in Figure S5) where the charges are injected by applying different gate voltages. 

The influence of the environment on the transport properties and electrical contacts of NWs limit their applications. Therefore, protecting NWs by preserving the electronic and magnetic properties of the isolated \ce{VS4} NW should be tested in spintronic applications. Nanotubes may protect NWs by preventing oxidation and maintaining electronic and magnetic properties unchanged\cite{pham2018torsional, li2018robust}, and nanocables, composed of insulating outer sheaths and a NW core, may guide the design of the models. Here, a designed nanocable on the isolated \ce{VS4} NW and a $1 \times 1 \times 5$ (8, 8) boron nitride (BN) zigzag nanotube is constructed (Figure S6), where the lattice mismatch is approximately 4.5\%. The isolated \ce{VS4} NW has a slight strain. The distance between the NW and the wall of the BN nanotube is approximately 3.57 \AA. The electronic and magnetic properties of the \ce{VS4-BN} nanocable is similar to that of the isolated \ce{VS4} NW. Therefore, the spin polarisation of the nanocable mainly derives from the inner NW, while the outer nanotube has a negligible effect on the nanocable. Such a hybrid structure may also enable interesting applications in spintronics.

\section{Summary}
We presented herein a new strategy in \ce{VS4} NWs towards AFM spintronics. The \ce{VS4} NWs are bound together by Van der Waals forces to form nano-rods in quasi-1D compounds in experiments. In this context, the geometric, electronic and magnetic properties of the \ce{VS4} bulk phase and isolated \ce{VS4} NW were analysed. First, we investigated the stability of the \ce{VS4} bulk phase and isolate \ce{VS4} NW by formation energy, AIMD and phonon spectra. 
After confirming the stability of \ce{VS4} NWs, we investigated the electronic and magnetic properties of \ce{VS4} NWs. The magnetic ground states of the isolate \ce{VS4} NW is AFM with a high N\'eel temperature (210 $K$). 
Detecting and manipulating the spin of AFM materials remains a major challenge due to spin degeneracy in the band structure. Nevertheless, carrier doping can separate the spin degeneracy to induce local spin polarisation. The isolated \ce{VS4} NW represent the half-metallic antiferromagnet resulting from carrier doping, which can be achieved with gate voltages. 
Our results indicate that the isolated \ce{VS4} NW is a promising 1D material for AFM spintronic applications. Carrier doping induces a rigid shift in the Fermi level into the valence or conduction bands, resulting in a complete spin-polarisation of carriers, which can be induced by applying a gate voltage. Thus, spin polarisation currents can be manipulated by applying a gate voltage, with a high potential for spintronic applications.
We further considered the protection of NWs. BN nanotubes can provide protection by preventing oxidation and by preserving the electronic and magnetic properties of \ce{VS4} NWs. Our results open up new oppportunities for applying 1D NWs in AFM spintronics by inducing half-metallic antiferromagnetism.

\section{Conflicts of interest}
There are no conflicts to declare.

\section{Acknowledgements}
Charles University Centre of Advanced Materials (CUCAM) (OP VVV Excellent Research Teams, Project No. CZ.02.1.01/0.0/0.0/15 \textunderscore 003/0000417) is acknowledged.
S. L. acknowledges the support from GAUK project (Grant No. 792218).
The results from this research have been achieved using the DECI resource Kay based in Poblacht na hÉireann at ICHEC with support from the PRACE DECI-16 16DECI0034 CABLE project.
We also acknowledge the computer resources, technical expertise, and assistance provided by IT4Innovations National Supercomputing Center. 
We thank Carlos V. Melo Carlos for editing the manuscript.

\bibliography{rsc} 

\providecommand*{\mcitethebibliography}{\thebibliography}
\csname @ifundefined\endcsname{endmcitethebibliography}
{\let\endmcitethebibliography\endthebibliography}{}
\begin{mcitethebibliography}{74}
\providecommand*{\natexlab}[1]{#1}
\providecommand*{\mciteSetBstSublistMode}[1]{}
\providecommand*{\mciteSetBstMaxWidthForm}[2]{}
\providecommand*{\mciteBstWouldAddEndPuncttrue}
  {\def\EndOfBibitem{\unskip.}}
\providecommand*{\mciteBstWouldAddEndPunctfalse}
  {\let\EndOfBibitem\relax}
\providecommand*{\mciteSetBstMidEndSepPunct}[3]{}
\providecommand*{\mciteSetBstSublistLabelBeginEnd}[3]{}
\providecommand*{\EndOfBibitem}{}
\mciteSetBstSublistMode{f}
\mciteSetBstMaxWidthForm{subitem}
{(\emph{\alph{mcitesubitemcount}})}
\mciteSetBstSublistLabelBeginEnd{\mcitemaxwidthsubitemform\space}
{\relax}{\relax}

\bibitem[Del~Alamo(2011)]{del2011nanometre}
J.~A. Del~Alamo, \emph{Nature}, 2011, \textbf{479}, 317\relax
\mciteBstWouldAddEndPuncttrue
\mciteSetBstMidEndSepPunct{\mcitedefaultmidpunct}
{\mcitedefaultendpunct}{\mcitedefaultseppunct}\relax
\EndOfBibitem
\bibitem[Tomioka \emph{et~al.}(2012)Tomioka, Yoshimura, and
  Fukui]{tomioka2012iii}
K.~Tomioka, M.~Yoshimura and T.~Fukui, \emph{Nature}, 2012, \textbf{488},
  189\relax
\mciteBstWouldAddEndPuncttrue
\mciteSetBstMidEndSepPunct{\mcitedefaultmidpunct}
{\mcitedefaultendpunct}{\mcitedefaultseppunct}\relax
\EndOfBibitem
\bibitem[Ferain \emph{et~al.}(2011)Ferain, Colinge, and
  Colinge]{ferain2011multigate}
I.~Ferain, C.~A. Colinge and J.-P. Colinge, \emph{Nature}, 2011, \textbf{479},
  310\relax
\mciteBstWouldAddEndPuncttrue
\mciteSetBstMidEndSepPunct{\mcitedefaultmidpunct}
{\mcitedefaultendpunct}{\mcitedefaultseppunct}\relax
\EndOfBibitem
\bibitem[Mack(2011)]{mack2011fifty}
C.~A. Mack, \emph{IEEE Transactions on Semiconductor Manufacturing}, 2011,
  \textbf{24}, 202--207\relax
\mciteBstWouldAddEndPuncttrue
\mciteSetBstMidEndSepPunct{\mcitedefaultmidpunct}
{\mcitedefaultendpunct}{\mcitedefaultseppunct}\relax
\EndOfBibitem
\bibitem[Lu and Lieber(2010)]{lu2010nanoelectronics}
W.~Lu and C.~M. Lieber, \emph{Nanoscience And Technology: A Collection of
  Reviews from Nature Journals}, World Scientific, 2010, pp. 137--146\relax
\mciteBstWouldAddEndPuncttrue
\mciteSetBstMidEndSepPunct{\mcitedefaultmidpunct}
{\mcitedefaultendpunct}{\mcitedefaultseppunct}\relax
\EndOfBibitem
\bibitem[Akinwande \emph{et~al.}(2014)Akinwande, Petrone, and
  Hone]{akinwande2014two}
D.~Akinwande, N.~Petrone and J.~Hone, \emph{Nature Communications}, 2014,
  \textbf{5}, 5678\relax
\mciteBstWouldAddEndPuncttrue
\mciteSetBstMidEndSepPunct{\mcitedefaultmidpunct}
{\mcitedefaultendpunct}{\mcitedefaultseppunct}\relax
\EndOfBibitem
\bibitem[Berger \emph{et~al.}(2004)Berger, Song, Li, Li, Ogbazghi, Feng, Dai,
  Marchenkov, Conrad, First,\emph{et~al.}]{berger2004ultrathin}
C.~Berger, Z.~Song, T.~Li, X.~Li, A.~Y. Ogbazghi, R.~Feng, Z.~Dai, A.~N.
  Marchenkov, E.~H. Conrad, P.~N. First \emph{et~al.}, \emph{The Journal of
  Physical Chemistry B}, 2004, \textbf{108}, 19912--19916\relax
\mciteBstWouldAddEndPuncttrue
\mciteSetBstMidEndSepPunct{\mcitedefaultmidpunct}
{\mcitedefaultendpunct}{\mcitedefaultseppunct}\relax
\EndOfBibitem
\bibitem[Wolf \emph{et~al.}(2001)Wolf, Awschalom, Buhrman, Daughton,
  Von~Molnar, Roukes, Chtchelkanova, and Treger]{wolf2001spintronics}
S.~Wolf, D.~Awschalom, R.~Buhrman, J.~Daughton, S.~Von~Molnar, M.~Roukes, A.~Y.
  Chtchelkanova and D.~Treger, \emph{science}, 2001, \textbf{294},
  1488--1495\relax
\mciteBstWouldAddEndPuncttrue
\mciteSetBstMidEndSepPunct{\mcitedefaultmidpunct}
{\mcitedefaultendpunct}{\mcitedefaultseppunct}\relax
\EndOfBibitem
\bibitem[Felser \emph{et~al.}(2007)Felser, Fecher, and
  Balke]{felser2007spintronics}
C.~Felser, G.~H. Fecher and B.~Balke, \emph{Angewandte Chemie International
  Edition}, 2007, \textbf{46}, 668--699\relax
\mciteBstWouldAddEndPuncttrue
\mciteSetBstMidEndSepPunct{\mcitedefaultmidpunct}
{\mcitedefaultendpunct}{\mcitedefaultseppunct}\relax
\EndOfBibitem
\bibitem[Li and Yang(2016)]{li2016first}
X.~Li and J.~Yang, \emph{National Science Review}, 2016, \textbf{3},
  365--381\relax
\mciteBstWouldAddEndPuncttrue
\mciteSetBstMidEndSepPunct{\mcitedefaultmidpunct}
{\mcitedefaultendpunct}{\mcitedefaultseppunct}\relax
\EndOfBibitem
\bibitem[Schaibley \emph{et~al.}(2016)Schaibley, Yu, Clark, Rivera, Ross,
  Seyler, Yao, and Xu]{schaibley2016valleytronics}
J.~R. Schaibley, H.~Yu, G.~Clark, P.~Rivera, J.~S. Ross, K.~L. Seyler, W.~Yao
  and X.~Xu, \emph{Nature Reviews Materials}, 2016, \textbf{1}, 16055\relax
\mciteBstWouldAddEndPuncttrue
\mciteSetBstMidEndSepPunct{\mcitedefaultmidpunct}
{\mcitedefaultendpunct}{\mcitedefaultseppunct}\relax
\EndOfBibitem
\bibitem[Zhong(2017)]{zhong2017d}
D.~Zhong, \emph{Science Advances}, 2017, \textbf{3}, e1603113\relax
\mciteBstWouldAddEndPuncttrue
\mciteSetBstMidEndSepPunct{\mcitedefaultmidpunct}
{\mcitedefaultendpunct}{\mcitedefaultseppunct}\relax
\EndOfBibitem
\bibitem[Jungwirth \emph{et~al.}(2018)Jungwirth, Sinova, Manchon, Marti,
  Wunderlich, and Felser]{jungwirth2018multiple}
T.~Jungwirth, J.~Sinova, A.~Manchon, X.~Marti, J.~Wunderlich and C.~Felser,
  \emph{Nature Physics}, 2018, \textbf{14}, 200--203\relax
\mciteBstWouldAddEndPuncttrue
\mciteSetBstMidEndSepPunct{\mcitedefaultmidpunct}
{\mcitedefaultendpunct}{\mcitedefaultseppunct}\relax
\EndOfBibitem
\bibitem[{\v{Z}}uti{\'c} \emph{et~al.}(2004){\v{Z}}uti{\'c}, Fabian, and
  Sarma]{vzutic2004spintronics}
I.~{\v{Z}}uti{\'c}, J.~Fabian and S.~D. Sarma, \emph{Reviews of Modern
  Physics}, 2004, \textbf{76}, 323\relax
\mciteBstWouldAddEndPuncttrue
\mciteSetBstMidEndSepPunct{\mcitedefaultmidpunct}
{\mcitedefaultendpunct}{\mcitedefaultseppunct}\relax
\EndOfBibitem
\bibitem[Chumak \emph{et~al.}(2015)Chumak, Vasyuchka, Serga, and
  Hillebrands]{chumak2015magnon}
A.~V. Chumak, V.~I. Vasyuchka, A.~A. Serga and B.~Hillebrands, \emph{Nature
  Physics}, 2015, \textbf{11}, 453--461\relax
\mciteBstWouldAddEndPuncttrue
\mciteSetBstMidEndSepPunct{\mcitedefaultmidpunct}
{\mcitedefaultendpunct}{\mcitedefaultseppunct}\relax
\EndOfBibitem
\bibitem[Camsari \emph{et~al.}(2015)Camsari, Ganguly, and
  Datta]{camsari2015modular}
K.~Y. Camsari, S.~Ganguly and S.~Datta, \emph{Scientific reports}, 2015,
  \textbf{5}, 10571\relax
\mciteBstWouldAddEndPuncttrue
\mciteSetBstMidEndSepPunct{\mcitedefaultmidpunct}
{\mcitedefaultendpunct}{\mcitedefaultseppunct}\relax
\EndOfBibitem
\bibitem[Mott(1936)]{mott1936resistance}
N.~F. Mott, \emph{Proceedings of the Royal Society of London. Series
  A-Mathematical and Physical Sciences}, 1936, \textbf{156}, 368--382\relax
\mciteBstWouldAddEndPuncttrue
\mciteSetBstMidEndSepPunct{\mcitedefaultmidpunct}
{\mcitedefaultendpunct}{\mcitedefaultseppunct}\relax
\EndOfBibitem
\bibitem[Picozzi(2014)]{picozzi2014ferroelectric}
S.~Picozzi, \emph{Frontiers in Physics}, 2014, \textbf{2}, 10\relax
\mciteBstWouldAddEndPuncttrue
\mciteSetBstMidEndSepPunct{\mcitedefaultmidpunct}
{\mcitedefaultendpunct}{\mcitedefaultseppunct}\relax
\EndOfBibitem
\bibitem[Mott(1936)]{mott1936electrical}
N.~F. Mott, \emph{Proceedings of the Royal Society of London. Series
  A-Mathematical and Physical Sciences}, 1936, \textbf{153}, 699--717\relax
\mciteBstWouldAddEndPuncttrue
\mciteSetBstMidEndSepPunct{\mcitedefaultmidpunct}
{\mcitedefaultendpunct}{\mcitedefaultseppunct}\relax
\EndOfBibitem
\bibitem[Li and Yang(2017)]{li2017low}
X.~Li and J.~Yang, \emph{Wiley Interdisciplinary Reviews: Computational
  Molecular Science}, 2017, \textbf{7}, e1314\relax
\mciteBstWouldAddEndPuncttrue
\mciteSetBstMidEndSepPunct{\mcitedefaultmidpunct}
{\mcitedefaultendpunct}{\mcitedefaultseppunct}\relax
\EndOfBibitem
\bibitem[Jungwirth \emph{et~al.}(2016)Jungwirth, Marti, Wadley, and
  Wunderlich]{jungwirth2016antiferromagnetic}
T.~Jungwirth, X.~Marti, P.~Wadley and J.~Wunderlich, \emph{Nature
  Nanotechnology}, 2016, \textbf{11}, 231\relax
\mciteBstWouldAddEndPuncttrue
\mciteSetBstMidEndSepPunct{\mcitedefaultmidpunct}
{\mcitedefaultendpunct}{\mcitedefaultseppunct}\relax
\EndOfBibitem
\bibitem[Baltz \emph{et~al.}(2018)Baltz, Manchon, Tsoi, Moriyama, Ono, and
  Tserkovnyak]{baltz2018antiferromagnetic}
V.~Baltz, A.~Manchon, M.~Tsoi, T.~Moriyama, T.~Ono and Y.~Tserkovnyak,
  \emph{Reviews of Modern Physics}, 2018, \textbf{90}, 015005\relax
\mciteBstWouldAddEndPuncttrue
\mciteSetBstMidEndSepPunct{\mcitedefaultmidpunct}
{\mcitedefaultendpunct}{\mcitedefaultseppunct}\relax
\EndOfBibitem
\bibitem[MacDonald and Tsoi(2011)]{macdonald2011antiferromagnetic}
A.~H. MacDonald and M.~Tsoi, \emph{Philosophical Transactions of the Royal
  Society A: Mathematical, Physical and Engineering Sciences}, 2011,
  \textbf{369}, 3098--3114\relax
\mciteBstWouldAddEndPuncttrue
\mciteSetBstMidEndSepPunct{\mcitedefaultmidpunct}
{\mcitedefaultendpunct}{\mcitedefaultseppunct}\relax
\EndOfBibitem
\bibitem[Hu(2012)]{hu2012half}
X.~Hu, \emph{Advanced Materials}, 2012, \textbf{24}, 294--298\relax
\mciteBstWouldAddEndPuncttrue
\mciteSetBstMidEndSepPunct{\mcitedefaultmidpunct}
{\mcitedefaultendpunct}{\mcitedefaultseppunct}\relax
\EndOfBibitem
\bibitem[He \emph{et~al.}(2015)He, Zhou, Jiao, Chen, Lu, and
  Sun]{he2015prediction}
J.~He, P.~Zhou, N.~Jiao, X.~Chen, W.~Lu and L.~Sun, \emph{RSC Advances}, 2015,
  \textbf{5}, 46640--46647\relax
\mciteBstWouldAddEndPuncttrue
\mciteSetBstMidEndSepPunct{\mcitedefaultmidpunct}
{\mcitedefaultendpunct}{\mcitedefaultseppunct}\relax
\EndOfBibitem
\bibitem[Nie and Hu(2008)]{nie2008possible}
Y.-m. Nie and X.~Hu, \emph{Physical Review Letters}, 2008, \textbf{100},
  117203\relax
\mciteBstWouldAddEndPuncttrue
\mciteSetBstMidEndSepPunct{\mcitedefaultmidpunct}
{\mcitedefaultendpunct}{\mcitedefaultseppunct}\relax
\EndOfBibitem
\bibitem[Van~Leuken and De~Groot(1995)]{van1995half}
H.~Van~Leuken and R.~De~Groot, \emph{Physical Review Letters}, 1995,
  \textbf{74}, 1171\relax
\mciteBstWouldAddEndPuncttrue
\mciteSetBstMidEndSepPunct{\mcitedefaultmidpunct}
{\mcitedefaultendpunct}{\mcitedefaultseppunct}\relax
\EndOfBibitem
\bibitem[Nayak \emph{et~al.}(2015)Nayak, Nicklas, Chadov, Khuntia, Shekhar,
  Kalache, Baenitz, Skourski, Guduru, Puri,\emph{et~al.}]{nayak2015design}
A.~K. Nayak, M.~Nicklas, S.~Chadov, P.~Khuntia, C.~Shekhar, A.~Kalache,
  M.~Baenitz, Y.~Skourski, V.~K. Guduru, A.~Puri \emph{et~al.}, \emph{Nature
  Materials}, 2015, \textbf{14}, 679\relax
\mciteBstWouldAddEndPuncttrue
\mciteSetBstMidEndSepPunct{\mcitedefaultmidpunct}
{\mcitedefaultendpunct}{\mcitedefaultseppunct}\relax
\EndOfBibitem
\bibitem[Kurt \emph{et~al.}(2014)Kurt, Rode, Stamenov, Venkatesan, Lau, Fonda,
  and Coey]{kurt2014cubic}
H.~Kurt, K.~Rode, P.~Stamenov, M.~Venkatesan, Y.-C. Lau, E.~Fonda and J.~Coey,
  \emph{Physical Review Letters}, 2014, \textbf{112}, 027201\relax
\mciteBstWouldAddEndPuncttrue
\mciteSetBstMidEndSepPunct{\mcitedefaultmidpunct}
{\mcitedefaultendpunct}{\mcitedefaultseppunct}\relax
\EndOfBibitem
\bibitem[Gong \emph{et~al.}(2018)Gong, Gong, Sun, Tong, Duan, Chu, and
  Zhang]{gong2018electrically}
S.-J. Gong, C.~Gong, Y.-Y. Sun, W.-Y. Tong, C.-G. Duan, J.-H. Chu and X.~Zhang,
  \emph{Proceedings of the National Academy of Sciences}, 2018, \textbf{115},
  8511--8516\relax
\mciteBstWouldAddEndPuncttrue
\mciteSetBstMidEndSepPunct{\mcitedefaultmidpunct}
{\mcitedefaultendpunct}{\mcitedefaultseppunct}\relax
\EndOfBibitem
\bibitem[Ai \emph{et~al.}(2018)Ai, Liu, Yang, Zhang, and Zhao]{ai2018two}
H.~Ai, X.~Liu, B.~Yang, X.~Zhang and M.~Zhao, \emph{The Journal of Physical
  Chemistry C}, 2018, \textbf{122}, 1846--1851\relax
\mciteBstWouldAddEndPuncttrue
\mciteSetBstMidEndSepPunct{\mcitedefaultmidpunct}
{\mcitedefaultendpunct}{\mcitedefaultseppunct}\relax
\EndOfBibitem
\bibitem[Chuang \emph{et~al.}(2015)Chuang, Ho, Smith, Sfigakis, Pepper, Chen,
  Fan, Griffiths, Farrer, Beere,\emph{et~al.}]{chuang2015all}
P.~Chuang, S.-C. Ho, L.~W. Smith, F.~Sfigakis, M.~Pepper, C.-H. Chen, J.-C.
  Fan, J.~Griffiths, I.~Farrer, H.~E. Beere \emph{et~al.}, \emph{Nature
  Nanotechnology}, 2015, \textbf{10}, 35--39\relax
\mciteBstWouldAddEndPuncttrue
\mciteSetBstMidEndSepPunct{\mcitedefaultmidpunct}
{\mcitedefaultendpunct}{\mcitedefaultseppunct}\relax
\EndOfBibitem
\bibitem[Deng \emph{et~al.}(2018)Deng, Yu, Song, Zhang, Wang, Sun, Yi, Wu, Wu,
  Zhu,\emph{et~al.}]{deng2018gate}
Y.~Deng, Y.~Yu, Y.~Song, J.~Zhang, N.~Z. Wang, Z.~Sun, Y.~Yi, Y.~Z. Wu, S.~Wu,
  J.~Zhu \emph{et~al.}, \emph{Nature}, 2018, \textbf{563}, 94--99\relax
\mciteBstWouldAddEndPuncttrue
\mciteSetBstMidEndSepPunct{\mcitedefaultmidpunct}
{\mcitedefaultendpunct}{\mcitedefaultseppunct}\relax
\EndOfBibitem
\bibitem[Li \emph{et~al.}(2014)Li, Wu, and Yang]{li2014half}
X.~Li, X.~Wu and J.~Yang, \emph{Journal of the American Chemical Society},
  2014, \textbf{136}, 11065--11069\relax
\mciteBstWouldAddEndPuncttrue
\mciteSetBstMidEndSepPunct{\mcitedefaultmidpunct}
{\mcitedefaultendpunct}{\mcitedefaultseppunct}\relax
\EndOfBibitem
\bibitem[Yuan \emph{et~al.}(2009)Yuan, Shimotani, Tsukazaki, Ohtomo, Kawasaki,
  and Iwasa]{yuan2009high}
H.~Yuan, H.~Shimotani, A.~Tsukazaki, A.~Ohtomo, M.~Kawasaki and Y.~Iwasa,
  \emph{Advanced Functional Materials}, 2009, \textbf{19}, 1046--1053\relax
\mciteBstWouldAddEndPuncttrue
\mciteSetBstMidEndSepPunct{\mcitedefaultmidpunct}
{\mcitedefaultendpunct}{\mcitedefaultseppunct}\relax
\EndOfBibitem
\bibitem[He \emph{et~al.}(2019)He, Ding, Zhong, Li, Li, and Zhang]{he2019cr}
J.~He, G.~Ding, C.~Zhong, S.~Li, D.~Li and G.~Zhang, \emph{Nanoscale}, 2019,
  \textbf{11}, 356--364\relax
\mciteBstWouldAddEndPuncttrue
\mciteSetBstMidEndSepPunct{\mcitedefaultmidpunct}
{\mcitedefaultendpunct}{\mcitedefaultseppunct}\relax
\EndOfBibitem
\bibitem[Pham \emph{et~al.}(2018)Pham, Oh, Stetz, Onishi, Kisielowski, Cohen,
  and Zettl]{pham2018torsional}
T.~Pham, S.~Oh, P.~Stetz, S.~Onishi, C.~Kisielowski, M.~L. Cohen and A.~Zettl,
  \emph{Science}, 2018, \textbf{361}, 263--266\relax
\mciteBstWouldAddEndPuncttrue
\mciteSetBstMidEndSepPunct{\mcitedefaultmidpunct}
{\mcitedefaultendpunct}{\mcitedefaultseppunct}\relax
\EndOfBibitem
\bibitem[Li \emph{et~al.}(2017)Li, Lv, Dai, Ma, Zeng, Wu, and Yang]{li2017half}
X.~Li, H.~Lv, J.~Dai, L.~Ma, X.~C. Zeng, X.~Wu and J.~Yang, \emph{Journal of
  the American Chemical Society}, 2017, \textbf{139}, 6290--6293\relax
\mciteBstWouldAddEndPuncttrue
\mciteSetBstMidEndSepPunct{\mcitedefaultmidpunct}
{\mcitedefaultendpunct}{\mcitedefaultseppunct}\relax
\EndOfBibitem
\bibitem[Wan \emph{et~al.}(2018)Wan, Sun, Wu, and Yang]{wan2018ambipolar}
Y.~Wan, Y.~Sun, X.~Wu and J.~Yang, \emph{The Journal of Physical Chemistry C},
  2018, \textbf{122}, 989--994\relax
\mciteBstWouldAddEndPuncttrue
\mciteSetBstMidEndSepPunct{\mcitedefaultmidpunct}
{\mcitedefaultendpunct}{\mcitedefaultseppunct}\relax
\EndOfBibitem
\bibitem[Zhang \emph{et~al.}(2016)Zhang, Zhu, and Chen]{zhang2016electron}
T.~Zhang, L.~Zhu and G.~Chen, \emph{Journal of Materials Chemistry C}, 2016,
  \textbf{4}, 10209--10214\relax
\mciteBstWouldAddEndPuncttrue
\mciteSetBstMidEndSepPunct{\mcitedefaultmidpunct}
{\mcitedefaultendpunct}{\mcitedefaultseppunct}\relax
\EndOfBibitem
\bibitem[Li \emph{et~al.}(2018)Li, Wang, Hu, Chen, Zhang, and
  Yan]{li2018robust}
S.-s. Li, Y.-p. Wang, S.-j. Hu, D.~Chen, C.-w. Zhang and S.-s. Yan,
  \emph{Nanoscale}, 2018, \textbf{10}, 15545--15552\relax
\mciteBstWouldAddEndPuncttrue
\mciteSetBstMidEndSepPunct{\mcitedefaultmidpunct}
{\mcitedefaultendpunct}{\mcitedefaultseppunct}\relax
\EndOfBibitem
\bibitem[Ye \emph{et~al.}(2017)Ye, Wang, Deng, Zeng, Nie, Duchesne, Wang, Liu,
  Zhou, Zhao,\emph{et~al.}]{ye2017amorphous}
H.~Ye, L.~Wang, S.~Deng, X.~Zeng, K.~Nie, P.~N. Duchesne, B.~Wang, S.~Liu,
  J.~Zhou, F.~Zhao \emph{et~al.}, \emph{Advanced Energy Materials}, 2017,
  \textbf{7}, 1601602\relax
\mciteBstWouldAddEndPuncttrue
\mciteSetBstMidEndSepPunct{\mcitedefaultmidpunct}
{\mcitedefaultendpunct}{\mcitedefaultseppunct}\relax
\EndOfBibitem
\bibitem[Shang \emph{et~al.}(2020)Shang, Fu, Zhou, and Zhao]{shang2020atomic}
C.~Shang, L.~Fu, S.~Zhou and J.~Zhao, \emph{JACS Au}, 2020,
  https://doi.org/10.1021/jacsau.0c00049\relax
\mciteBstWouldAddEndPuncttrue
\mciteSetBstMidEndSepPunct{\mcitedefaultmidpunct}
{\mcitedefaultendpunct}{\mcitedefaultseppunct}\relax
\EndOfBibitem
\bibitem[Hillebrand(1907)]{hillebrand1907vanadium}
W.~Hillebrand, \emph{Journal of the American Chemical Society}, 1907,
  \textbf{29}, 1019--1029\relax
\mciteBstWouldAddEndPuncttrue
\mciteSetBstMidEndSepPunct{\mcitedefaultmidpunct}
{\mcitedefaultendpunct}{\mcitedefaultseppunct}\relax
\EndOfBibitem
\bibitem[Flores \emph{et~al.}(2018)Flores, Munoz-Cortes, Bodega,
  Caballero-Calero, Martin-Gonzaalez, Sanchez, Ares, and
  Ferrer]{flores2018beyond}
E.~Flores, E.~Munoz-Cortes, J.~Bodega, O.~Caballero-Calero,
  M.~Martin-Gonzaalez, C.~Sanchez, J.~R. Ares and I.~J. Ferrer, \emph{ACS
  Applied Energy Materials}, 2018, \textbf{1}, 2333--2340\relax
\mciteBstWouldAddEndPuncttrue
\mciteSetBstMidEndSepPunct{\mcitedefaultmidpunct}
{\mcitedefaultendpunct}{\mcitedefaultseppunct}\relax
\EndOfBibitem
\bibitem[Zhou \emph{et~al.}(2016)Zhou, Li, Yang, Tian, Xu, Yang, and
  Fan]{zhou2016conductive}
Y.~Zhou, Y.~Li, J.~Yang, J.~Tian, H.~Xu, J.~Yang and W.~Fan, \emph{ACS Applied
  Materials \& Interfaces}, 2016, \textbf{8}, 18797--18805\relax
\mciteBstWouldAddEndPuncttrue
\mciteSetBstMidEndSepPunct{\mcitedefaultmidpunct}
{\mcitedefaultendpunct}{\mcitedefaultseppunct}\relax
\EndOfBibitem
\bibitem[Lui \emph{et~al.}(2015)Lui, Jiang, Duan, Broughton, Zhang, Fowler, and
  Yu]{lui2015synthesis}
G.~Lui, G.~Jiang, A.~Duan, J.~Broughton, J.~Zhang, M.~W. Fowler and A.~Yu,
  \emph{Industrial \& Engineering Chemistry Research}, 2015, \textbf{54},
  2682--2689\relax
\mciteBstWouldAddEndPuncttrue
\mciteSetBstMidEndSepPunct{\mcitedefaultmidpunct}
{\mcitedefaultendpunct}{\mcitedefaultseppunct}\relax
\EndOfBibitem
\bibitem[Rout \emph{et~al.}(2013)Rout, Kim, Xu, Yang, Jeong, Odkhuu, Park, Cho,
  and Shin]{rout2013synthesis}
C.~S. Rout, B.-H. Kim, X.~Xu, J.~Yang, H.~Y. Jeong, D.~Odkhuu, N.~Park, J.~Cho
  and H.~S. Shin, \emph{Journal of the American Chemical Society}, 2013,
  \textbf{135}, 8720--8725\relax
\mciteBstWouldAddEndPuncttrue
\mciteSetBstMidEndSepPunct{\mcitedefaultmidpunct}
{\mcitedefaultendpunct}{\mcitedefaultseppunct}\relax
\EndOfBibitem
\bibitem[Sun \emph{et~al.}(2015)Sun, Wei, Li, Luo, An, Sheng, Wang, Chen, and
  Mai]{sun2015vanadium}
R.~Sun, Q.~Wei, Q.~Li, W.~Luo, Q.~An, J.~Sheng, D.~Wang, W.~Chen and L.~Mai,
  \emph{ACS Applied Materials \& Interfaces}, 2015, \textbf{7},
  20902--20908\relax
\mciteBstWouldAddEndPuncttrue
\mciteSetBstMidEndSepPunct{\mcitedefaultmidpunct}
{\mcitedefaultendpunct}{\mcitedefaultseppunct}\relax
\EndOfBibitem
\bibitem[Wang \emph{et~al.}(2018)Wang, Gong, Yang, Liao, Wu, Xu, Chen, Yang,
  Zhao, Wang,\emph{et~al.}]{wang2018graphene}
S.~Wang, F.~Gong, S.~Yang, J.~Liao, M.~Wu, Z.~Xu, C.~Chen, X.~Yang, F.~Zhao,
  B.~Wang \emph{et~al.}, \emph{Advanced Functional Materials}, 2018,
  \textbf{28}, 1801806\relax
\mciteBstWouldAddEndPuncttrue
\mciteSetBstMidEndSepPunct{\mcitedefaultmidpunct}
{\mcitedefaultendpunct}{\mcitedefaultseppunct}\relax
\EndOfBibitem
\bibitem[Wang \emph{et~al.}(2018)Wang, Liu, Wang, Yi, Chen, Ma, Hu, Zhu, Chen,
  Tie,\emph{et~al.}]{wang2018highly}
Y.~Wang, Z.~Liu, C.~Wang, X.~Yi, R.~Chen, L.~Ma, Y.~Hu, G.~Zhu, T.~Chen, Z.~Tie
  \emph{et~al.}, \emph{Advanced Materials}, 2018, \textbf{30}, 1802563\relax
\mciteBstWouldAddEndPuncttrue
\mciteSetBstMidEndSepPunct{\mcitedefaultmidpunct}
{\mcitedefaultendpunct}{\mcitedefaultseppunct}\relax
\EndOfBibitem
\bibitem[Gao \emph{et~al.}(2016)Gao, Ding, Li, Yao, Wu, and
  Qian]{gao2016monolayer}
G.~Gao, G.~Ding, J.~Li, K.~Yao, M.~Wu and M.~Qian, \emph{Nanoscale}, 2016,
  \textbf{8}, 8986--8994\relax
\mciteBstWouldAddEndPuncttrue
\mciteSetBstMidEndSepPunct{\mcitedefaultmidpunct}
{\mcitedefaultendpunct}{\mcitedefaultseppunct}\relax
\EndOfBibitem
\bibitem[Frey \emph{et~al.}(2019)Frey, Bandyopadhyay, Kumar, Anasori, Gogotsi,
  and Shenoy]{frey2019surface}
N.~C. Frey, A.~Bandyopadhyay, H.~Kumar, B.~Anasori, Y.~Gogotsi and V.~B.
  Shenoy, \emph{ACS nano}, 2019, \textbf{13}, 2831--2839\relax
\mciteBstWouldAddEndPuncttrue
\mciteSetBstMidEndSepPunct{\mcitedefaultmidpunct}
{\mcitedefaultendpunct}{\mcitedefaultseppunct}\relax
\EndOfBibitem
\bibitem[Ma \emph{et~al.}(2017)Ma, Kuc, Jing, Philipsen, and Heine]{ma2017two}
Y.~Ma, A.~Kuc, Y.~Jing, P.~Philipsen and T.~Heine, \emph{Angewandte Chemie
  International Edition}, 2017, \textbf{56}, 10214--10218\relax
\mciteBstWouldAddEndPuncttrue
\mciteSetBstMidEndSepPunct{\mcitedefaultmidpunct}
{\mcitedefaultendpunct}{\mcitedefaultseppunct}\relax
\EndOfBibitem
\bibitem[Kresse(1993)]{kresse1993g}
G.~Kresse, \emph{Physical Review B}, 1993, \textbf{47}, 558\relax
\mciteBstWouldAddEndPuncttrue
\mciteSetBstMidEndSepPunct{\mcitedefaultmidpunct}
{\mcitedefaultendpunct}{\mcitedefaultseppunct}\relax
\EndOfBibitem
\bibitem[Kresse(1999)]{kresse1999g}
G.~Kresse, \emph{Physical Review B}, 1999, \textbf{59}, 1758\relax
\mciteBstWouldAddEndPuncttrue
\mciteSetBstMidEndSepPunct{\mcitedefaultmidpunct}
{\mcitedefaultendpunct}{\mcitedefaultseppunct}\relax
\EndOfBibitem
\bibitem[Perdew \emph{et~al.}(1996)Perdew, Burke, and
  Ernzerhof]{perdew1996generalized}
J.~P. Perdew, K.~Burke and M.~Ernzerhof, \emph{Physical Review Letters}, 1996,
  \textbf{77}, 3865\relax
\mciteBstWouldAddEndPuncttrue
\mciteSetBstMidEndSepPunct{\mcitedefaultmidpunct}
{\mcitedefaultendpunct}{\mcitedefaultseppunct}\relax
\EndOfBibitem
\bibitem[Heyd \emph{et~al.}(2003)Heyd, Scuseria, and Ernzerhof]{heyd2003hybrid}
J.~Heyd, G.~E. Scuseria and M.~Ernzerhof, \emph{The Journal of Chemical
  Physics}, 2003, \textbf{118}, 8207--8215\relax
\mciteBstWouldAddEndPuncttrue
\mciteSetBstMidEndSepPunct{\mcitedefaultmidpunct}
{\mcitedefaultendpunct}{\mcitedefaultseppunct}\relax
\EndOfBibitem
\bibitem[Grimme \emph{et~al.}(2010)Grimme, Antony, Ehrlich, and
  Krieg]{grimme2010consistent}
S.~Grimme, J.~Antony, S.~Ehrlich and H.~Krieg, \emph{The Journal of Chemical
  Physics}, 2010, \textbf{132}, 154104\relax
\mciteBstWouldAddEndPuncttrue
\mciteSetBstMidEndSepPunct{\mcitedefaultmidpunct}
{\mcitedefaultendpunct}{\mcitedefaultseppunct}\relax
\EndOfBibitem
\bibitem[Togo and Tanaka(2015)]{togo2015phonopy}
A.~Togo and I.~Tanaka, \emph{Scr. Mater.}, 2015, \textbf{108}, 1--5\relax
\mciteBstWouldAddEndPuncttrue
\mciteSetBstMidEndSepPunct{\mcitedefaultmidpunct}
{\mcitedefaultendpunct}{\mcitedefaultseppunct}\relax
\EndOfBibitem
\bibitem[Nos{\'e}(1984)]{nose1984unified}
S.~Nos{\'e}, \emph{The Journal of Chemical Physics}, 1984, \textbf{81},
  511--519\relax
\mciteBstWouldAddEndPuncttrue
\mciteSetBstMidEndSepPunct{\mcitedefaultmidpunct}
{\mcitedefaultendpunct}{\mcitedefaultseppunct}\relax
\EndOfBibitem
\bibitem[Watanabe(1974)]{watanabe1974crystal}
Y.~Watanabe, \emph{Acta Crystallographica Section B: Structural Crystallography
  and Crystal Chemistry}, 1974, \textbf{30}, 1396--1401\relax
\mciteBstWouldAddEndPuncttrue
\mciteSetBstMidEndSepPunct{\mcitedefaultmidpunct}
{\mcitedefaultendpunct}{\mcitedefaultseppunct}\relax
\EndOfBibitem
\bibitem[McCoy and Wu(2014)]{mccoy2014two}
B.~M. McCoy and T.~T. Wu, \emph{The two-dimensional Ising model}, Courier
  Corporation, 2014\relax
\mciteBstWouldAddEndPuncttrue
\mciteSetBstMidEndSepPunct{\mcitedefaultmidpunct}
{\mcitedefaultendpunct}{\mcitedefaultseppunct}\relax
\EndOfBibitem
\bibitem[Albuquerque \emph{et~al.}(2007)Albuquerque, Alet, Corboz, Dayal,
  Feiguin, Fuchs, Gamper, Gull, G{\"u}rtler,
  Honecker,\emph{et~al.}]{albuquerque2007alps}
A.~F. Albuquerque, F.~Alet, P.~Corboz, P.~Dayal, A.~Feiguin, S.~Fuchs,
  L.~Gamper, E.~Gull, S.~G{\"u}rtler, A.~Honecker \emph{et~al.}, \emph{Journal
  of Magnetism and Magnetic Materials}, 2007, \textbf{310}, 1187--1193\relax
\mciteBstWouldAddEndPuncttrue
\mciteSetBstMidEndSepPunct{\mcitedefaultmidpunct}
{\mcitedefaultendpunct}{\mcitedefaultseppunct}\relax
\EndOfBibitem
\bibitem[Zhou \emph{et~al.}(2018)Zhou, Zang, Wei, Zheng, Hao, Ling, Tang, Fang,
  and Zhou]{ZHOU2018118}
T.~Zhou, Z.~Zang, J.~Wei, J.~Zheng, J.~Hao, F.~Ling, X.~Tang, L.~Fang and
  M.~Zhou, \emph{Nano Energy}, 2018, \textbf{50}, 118 -- 125\relax
\mciteBstWouldAddEndPuncttrue
\mciteSetBstMidEndSepPunct{\mcitedefaultmidpunct}
{\mcitedefaultendpunct}{\mcitedefaultseppunct}\relax
\EndOfBibitem
\bibitem[{Ur Rehman} \emph{et~al.}(2017){Ur Rehman}, Li, Li, and
  Ding]{URREHMAN2017163}
S.~{Ur Rehman}, Z.~Li, H.~Li and Z.~Ding, \emph{Physica B: Condensed Matter},
  2017, \textbf{524}, 163 -- 172\relax
\mciteBstWouldAddEndPuncttrue
\mciteSetBstMidEndSepPunct{\mcitedefaultmidpunct}
{\mcitedefaultendpunct}{\mcitedefaultseppunct}\relax
\EndOfBibitem
\bibitem[Lou and Lee(2019)]{PingOrigin}
P.~Lou and J.~Y. Lee, \emph{The Journal of Chemical Physics}, 2019,
  \textbf{150}, 184307\relax
\mciteBstWouldAddEndPuncttrue
\mciteSetBstMidEndSepPunct{\mcitedefaultmidpunct}
{\mcitedefaultendpunct}{\mcitedefaultseppunct}\relax
\EndOfBibitem
\bibitem[Hotta \emph{et~al.}(1999)Hotta, Yunoki, Mayr, and
  Dagotto]{takashiAtype}
T.~Hotta, S.~Yunoki, M.~Mayr and E.~Dagotto, \emph{Physical Review B}, 1999,
  \textbf{60}, 15009--15012\relax
\mciteBstWouldAddEndPuncttrue
\mciteSetBstMidEndSepPunct{\mcitedefaultmidpunct}
{\mcitedefaultendpunct}{\mcitedefaultseppunct}\relax
\EndOfBibitem
\bibitem[Gambardella(2008)]{Gambardellamagnetism}
P.~Gambardella, Magnetic Nanostructures in Modern Technology, Dordrecht, 2008,
  pp. 325--342\relax
\mciteBstWouldAddEndPuncttrue
\mciteSetBstMidEndSepPunct{\mcitedefaultmidpunct}
{\mcitedefaultendpunct}{\mcitedefaultseppunct}\relax
\EndOfBibitem
\bibitem[Takahashi \emph{et~al.}(1991)Takahashi, Turek, Nakazawa, Tamura,
  Nozawa, Shiomi, Ishikawa, and Kinoshita]{takahashi1991discovery}
M.~Takahashi, P.~Turek, Y.~Nakazawa, M.~Tamura, K.~Nozawa, D.~Shiomi,
  M.~Ishikawa and M.~Kinoshita, \emph{Physical Review Letters}, 1991,
  \textbf{67}, 746\relax
\mciteBstWouldAddEndPuncttrue
\mciteSetBstMidEndSepPunct{\mcitedefaultmidpunct}
{\mcitedefaultendpunct}{\mcitedefaultseppunct}\relax
\EndOfBibitem
\bibitem[Tan \emph{et~al.}(2020)Tan, Liu, Xiang, Du, Lou, and Fu]{tan2020one}
X.~Tan, L.~Liu, H.~Xiang, G.-F. Du, A.~Lou and H.-H. Fu, \emph{Nanoscale},
  2020, \textbf{12}, 8942--8948\relax
\mciteBstWouldAddEndPuncttrue
\mciteSetBstMidEndSepPunct{\mcitedefaultmidpunct}
{\mcitedefaultendpunct}{\mcitedefaultseppunct}\relax
\EndOfBibitem
\bibitem[Huang \emph{et~al.}(2018)Huang, Clark, Klein, MacNeill,
  Navarro-Moratalla, Seyler, Wilson, McGuire, Cobden,
  Xiao,\emph{et~al.}]{huang2018electrical}
B.~Huang, G.~Clark, D.~R. Klein, D.~MacNeill, E.~Navarro-Moratalla, K.~L.
  Seyler, N.~Wilson, M.~A. McGuire, D.~H. Cobden, D.~Xiao \emph{et~al.},
  \emph{Nature Nanotechnology}, 2018, \textbf{13}, 544--548\relax
\mciteBstWouldAddEndPuncttrue
\mciteSetBstMidEndSepPunct{\mcitedefaultmidpunct}
{\mcitedefaultendpunct}{\mcitedefaultseppunct}\relax
\EndOfBibitem
\bibitem[Yang \emph{et~al.}(2018)Yang, Liu, Xu, Meng, Tong, Ma, Zhou, Sun, and
  Sheng]{yang2018antiferromagnetism}
Y.~Yang, C.~Liu, X.~Xu, Z.~Meng, W.~Tong, Z.~Ma, C.~Zhou, Y.~Sun and Z.~Sheng,
  \emph{Polymer Chemistry}, 2018, \textbf{9}, 5499--5503\relax
\mciteBstWouldAddEndPuncttrue
\mciteSetBstMidEndSepPunct{\mcitedefaultmidpunct}
{\mcitedefaultendpunct}{\mcitedefaultseppunct}\relax
\EndOfBibitem
\bibitem[Ueno \emph{et~al.}(2011)Ueno, Nakamura, Shimotani, Yuan, Kimura,
  Nojima, Aoki, Iwasa, and Kawasaki]{ueno2011discovery}
K.~Ueno, S.~Nakamura, H.~Shimotani, H.~Yuan, N.~Kimura, T.~Nojima, H.~Aoki,
  Y.~Iwasa and M.~Kawasaki, \emph{Nature Nanotechnology}, 2011, \textbf{6},
  408\relax
\mciteBstWouldAddEndPuncttrue
\mciteSetBstMidEndSepPunct{\mcitedefaultmidpunct}
{\mcitedefaultendpunct}{\mcitedefaultseppunct}\relax
\EndOfBibitem
\end{mcitethebibliography}
\bibliographystyle{rsc} 

\end{document}


\title{Doping isolated one-dimensional antiferromagnetic semiconductor Vanadium tetrasulfide (\ce{VS4}) nanowires with carriers induces half-metallicity}

\author{Shuo Li \textit{$^{a}$}, Junjie He \textit{$^{a,b}$}, Petr Nachtigall \textit{$^{a}$}, Luk\'a\v s Grajciar \textit{$^{a}$},  Federico Brivio \textit{$^{a}$}}
\maketitle

\footnotetext{\textit{$^{a}$Department of Physical and Macromolecular Chemistry \& Charles University Center of Advanced Materials, Faculty of Science, Charles University, Hlavova 8, 128 43 Prague 2, Czech Republic; E-mail: briviof@natur.cuni.cz}}
\footnotetext{\textit{$^{b}$Bremen Center for Computational Materials Science, University of Bremen, Am Fallturm 1, 28359 Bremen, Germany}}
\section{Figures}

\begin{figure}[h!]
\centering
\includegraphics[scale=0.5]{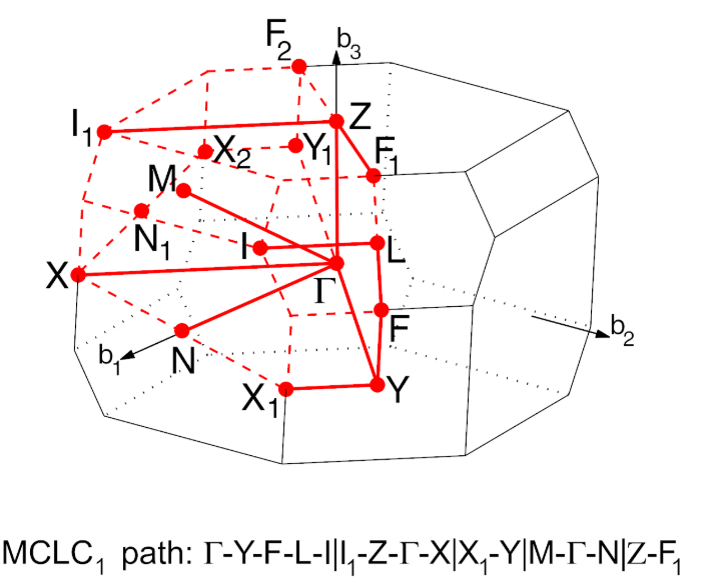}
\caption{The high symmetry points of the first Brillouin zone for the band structure of the \ce{VS4} bulk phase. This figure was created from http://materials.duke.edu/awrapper.html} 
\label{fig:S1}
\end{figure}

\begin{figure}[h!]
\centering
\includegraphics[scale=0.5]{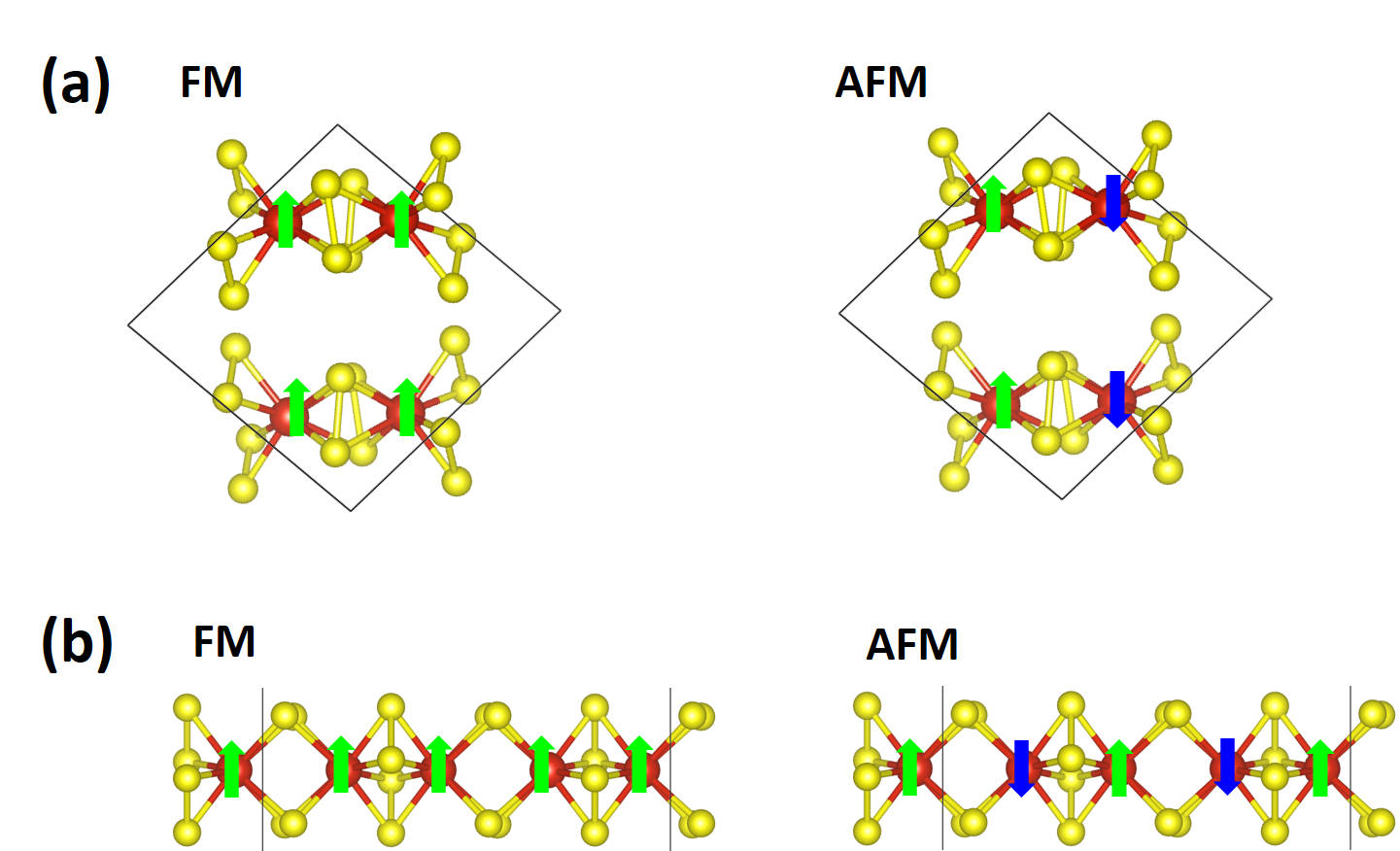}
\caption{Ferromagnetic (FM) and antiferromagnetic (AFM) states for (a) \ce{VS4} bulk phases and (b) isolated \ce{VS4} nanowires (NWs).}
\label{fig:S2}
\end{figure}

\begin{figure}[h!]
\centering
\includegraphics[scale=0.5]{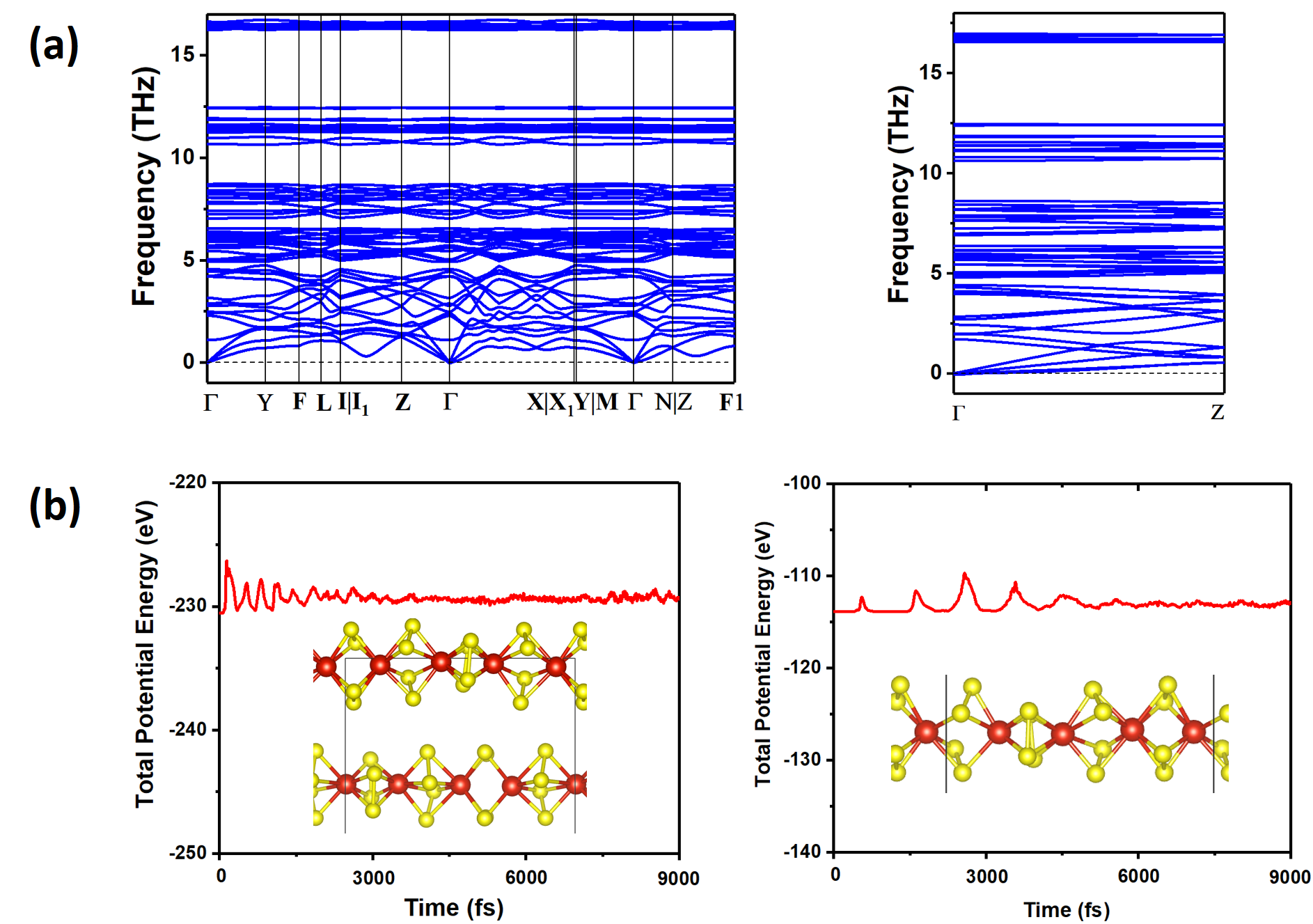}
\caption{(a) Phonon spectra and (b) total potential energy of NWs as a function of simulation time for the \ce{VS4} bulk phase (conventional cell) and the isolated \ce{VS4} NW by using \textit{ab-initio} molecular dynamics (300$K$). The inset shows the corresponding structure after the simulation for 9 ps.}
\label{fig:S3}
\end{figure}

\clearpage
\ce{T_N} is defined as a maximum on the temperature dependent specific heat \ce{C_V} curve, 
 \begin{equation}
C_V = (<E^2> - <E>^2)/T^2
\end{equation}
Where $T$ is the temperature and $E$ is the energy.

\begin{figure}[h!]
\centering
\includegraphics[scale=0.4]{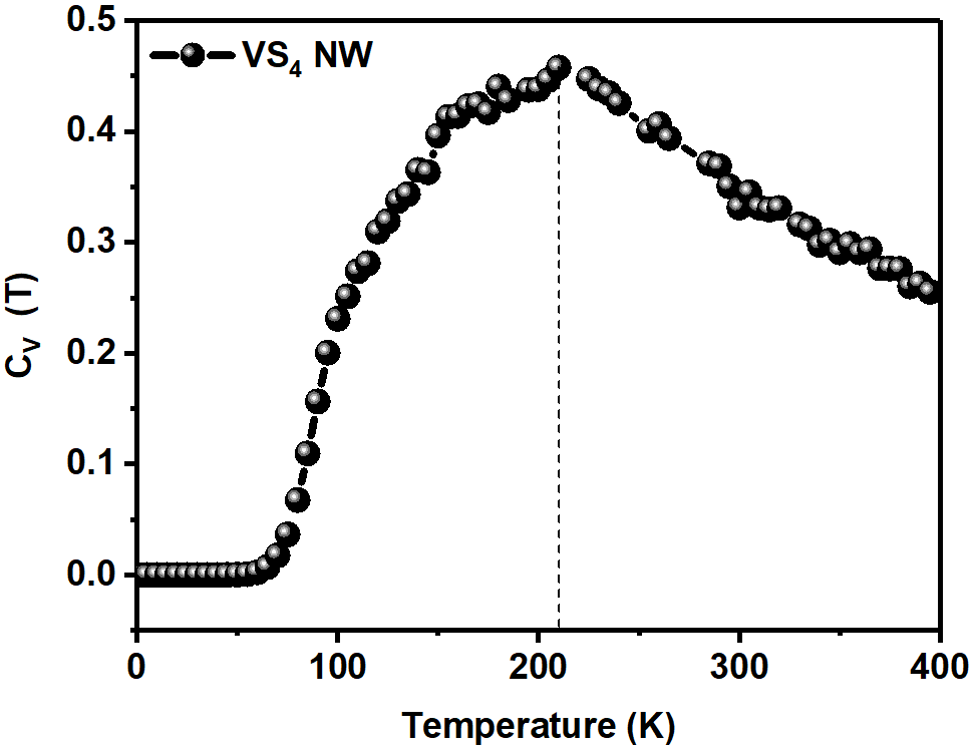}
\caption{Specific heat calculated for the isolated \ce{VS4} NW with respect to the temperature}
\label{fig:S4}
\end{figure}

\begin{figure}[h!]  
\centering
  \includegraphics[scale=0.4]{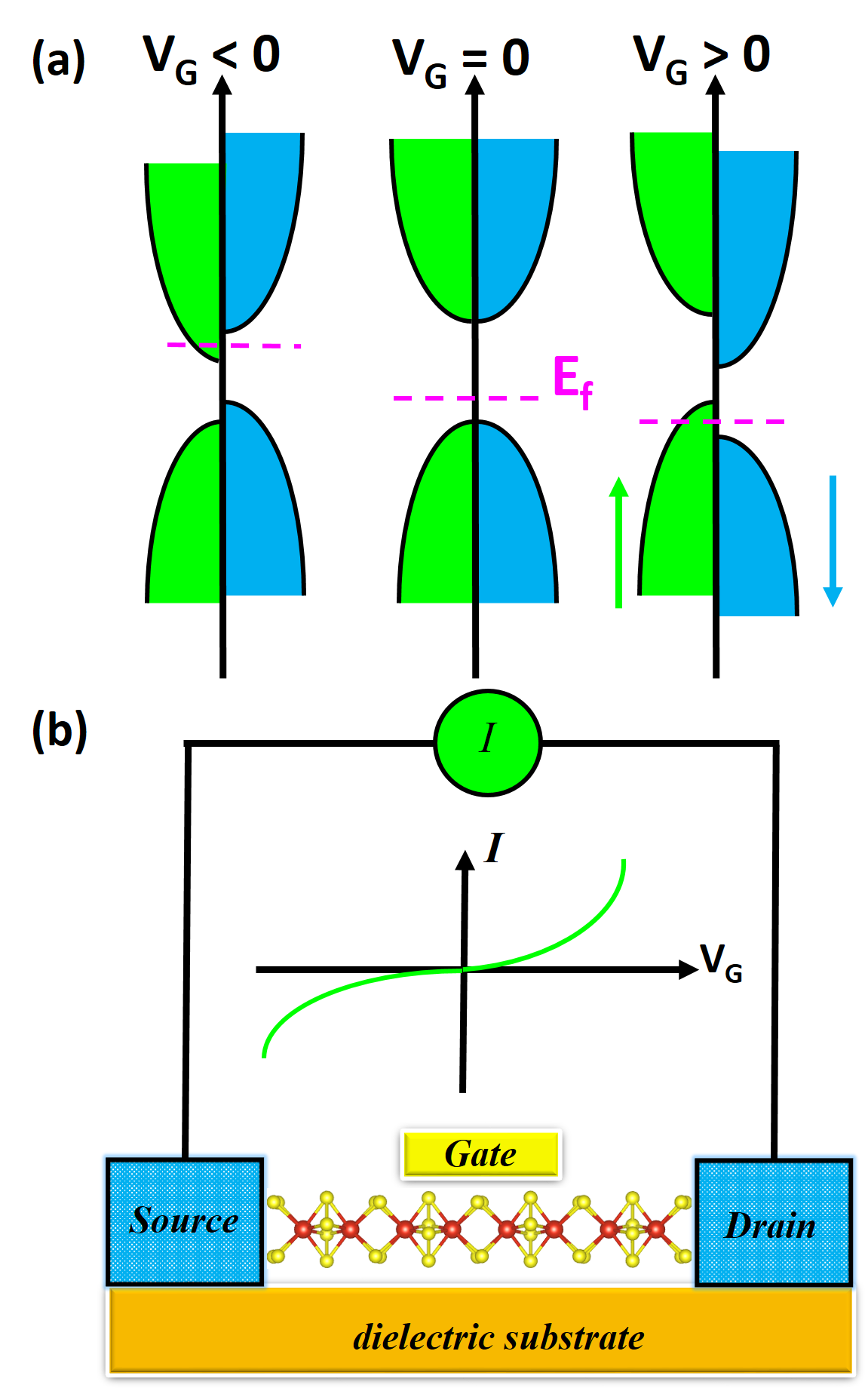} 
  \caption{(a) The schematic plot of transformation between the HMAF and AFM states under gate voltages of \ce{V_G} $<$ 0, \ce{V_G} = 0 and \ce{V_G} $>$ 0, respectively. (b) The schematic plot of AFM spintronics device based on the isolated \ce{VS4} NW, together with the \ce{{I-V}_G} relationship under the gate voltage. The switching of the spin current can be manipulated by gate voltages.}
  \label{fig:S5}
\end{figure}

\clearpage
To evaluate the stability of the isolated \ce{VS4} NW inside BN nanotubes of different sizes, the binding energy (\ce{E_b}) for the unit cell is calculated as:
\begin{equation}
E_{b} = E_{total} - E_{NW}- E_{nanotube}
\end{equation}
where \ce{E_{total}} stand for the total energies of the isolated \ce{VS4} NW inside the ($m$, $m$) BN nanotube. \ce{E_{NW}} and \ce{E_{nanotube}} are the energies of the isolated \ce{VS4} NW and the ($m$, $m$) BN nanotube. $m$ = 6, 7, 8 and 9, which are the direction of the vector.

\begin{figure}[h!]
\centering
\includegraphics[scale=0.5]{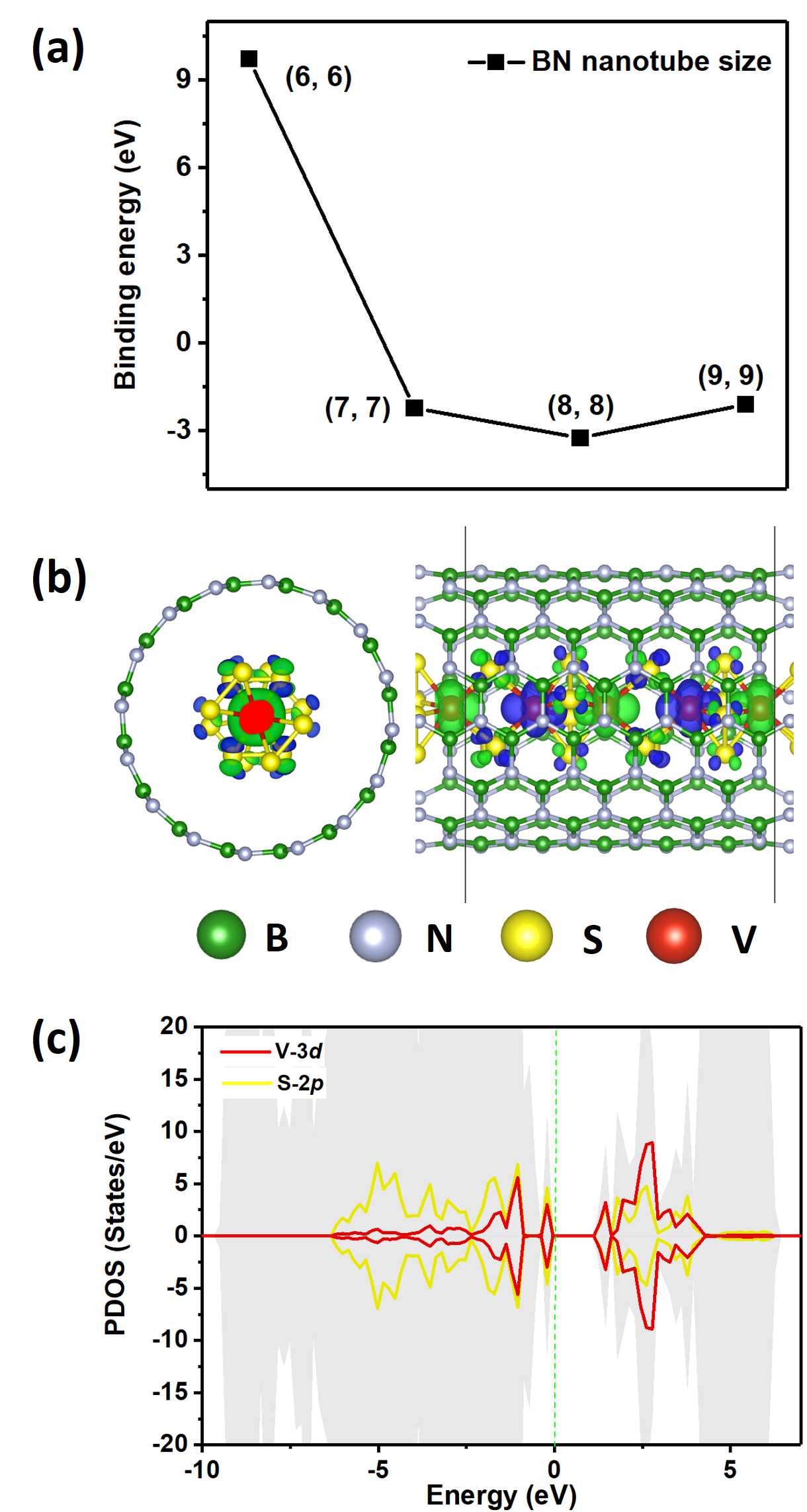}
\caption{(a) Binding energies \ce{E_b} of the isolated \ce{VS4} NW inside various BN nanotubes are presented. (b) Spin polarization density of the nanocable. (c) The partial density of states of the nanocable at $PBE+U$ level ($U=3$). The gray shadow is the total density of states.}
\label{fig:S6}
\end{figure}

